\documentclass[a4paper,11pt,twoside]{article}
\usepackage[paperheight=29.7cm,paperwidth=21cm,outer=2.25cm,inner=2.25cm,top=2cm,bottom=2cm,headheight=13.6pt,includehead,includefoot]{geometry} 
\usepackage{silence}
\newcommand{\textlineskip}{\baselineskip=13pt}

\providecommand{\keywords}[1]{\noindent\textbf{Keywords:} #1}

\def\fnt#1#2{\footnotetext{\kern-.3em%
          {$^{\mbox{\scriptsize #1}}$}{#2}}}

\usepackage{fancyhdr}

\fancypagestyle{pagestyle}{%
  \fancyhf{}
  \fancyhead[OR]{E. Tolotti, A. Jnini, F. Vella, R. Passerone}
  \fancyhead[EL]{cuVegas: CUDA-based Implementation of the VEGAS Enhanced Algorithm}
  \fancyfoot[C]{\thepage}%
}

\fancypagestyle{firstpagestyle}{%
  \fancyhf{}
  
  \fancyfoot[C]{\thepage}
}

\renewcommand{\thefootnote}{\fnsymbol{footnote}}  

\usepackage{amsmath}
\usepackage{amssymb}

\usepackage{bm}

\usepackage[figuresright]{rotating}

\usepackage[hyphens]{url}
\usepackage[hyperfootnotes=false]{hyperref}
\hypersetup{
  colorlinks,
  linkcolor={black},
  citecolor={black},
  urlcolor={black}
}
\usepackage[noabbrev,capitalise]{cleveref}

\usepackage{csquotes}
\usepackage[
backend=biber,
style=numeric,
sorting=none,
]{biblatex}

\usepackage{soul}

\usepackage{etoolbox}
\usepackage{accents}
\robustify{\underaccent}

\usepackage[ruled]{algorithm2e}
\SetAlFnt{\small}

\usepackage{multirow}
\usepackage{todonotes}
\usepackage{caption}
\usepackage{subcaption}
\usepackage{xcolor} 
\usepackage{tikz}
\usetikzlibrary{fit,calc}
\usepackage{colortbl}
\usepackage{booktabs} 

\usepackage{lmodern}

\let\oldtable\table
\let\endoldtable\endtable

\renewenvironment{table}[1][]{
  \oldtable[#1]
  \centering
  \smaller
}{\endoldtable}

\begin{document}

\newcommand{\roby}[1]{\todo[color=green]{{\footnotesize roby: #1}}}

\newcommand*{\tikzmk}[1]{\tikz[remember picture,overlay,] \node (#1) {};\ignorespaces}
\newcommand{\boxit}[1]{\tikz[remember picture,overlay]{\node[yshift=3pt,fill=#1,opacity=.25,fit={(A)($(B)+(.95\linewidth,.8\baselineskip)$)}] {};}\ignorespaces}
\colorlet{myorange}{orange!50}
\colorlet{myred}{red!50}
\colorlet{myblue}{cyan!50}
\colorlet{mygreen}{green!50}
\colorlet{myyellow}{yellow!50}

\newcommand{\bboxit}[2]{
    \tikz[remember picture,overlay] \node (A) {};\ignorespaces
    \tikz[remember picture,overlay]{\node[yshift=3pt,fill=#1,opacity=.25,fit={($(A)+(0,0.15\baselineskip)$)($(A)+(.95\linewidth,-{#2}\baselineskip - 0.25\baselineskip)$)}] {};}\ignorespaces
}
\newcommand{\bboxitt}[2]{
    \tikz[remember picture,overlay] \node (A) {};\ignorespaces
    \tikz[remember picture,overlay]{\node[yshift=3pt,fill=#1,opacity=.25,fit={($(A)+(0,0.15\baselineskip)$)($(A)+(.75\linewidth,-{#2}\baselineskip - 0.25\baselineskip)$)}] {};}\ignorespaces
}

\def\HilR{\leavevmode\rlap{\hbox to \hsize{{myred}\leaders\hrule height .9\baselineskip depth .5ex\hfill}}}

\def\HilB{\leavevmode\rlap{\hbox to \hsize{{myblue}\leaders\hrule height .9\baselineskip depth .5ex\hfill}}}

\def\HilG{\leavevmode\rlap{\hbox to \hsize{{mygreen}\leaders\hrule height .9\baselineskip depth .5ex\hfill}}}

\def\HilY{\leavevmode\rlap{\hbox to \hsize{{myyellow}\leaders\hrule height .9\baselineskip depth .5ex\hfill}}}

\normalsize\textlineskip
\thispagestyle{firstpagestyle}
\setcounter{page}{1}

\title{\vspace{-2cm}\Large\bf cuVegas: Accelerate Multidimensional Monte Carlo Integration through a Parallelized CUDA-based Implementation of the VEGAS Enhanced Algorithm}

\author{
Emiliano Tolotti\textsuperscript{1}\\
\texttt{\small{emiliano.tolotti@unitn.it}}
\and
Anas Jnini\textsuperscript{1}\\
\texttt{\small{anas.jnini@unitn.it}}
\and
Flavio Vella\textsuperscript{1}\\
\texttt{\small{flavio.vella@unitn.it}}
\and
Roberto Passerone\textsuperscript{1}\\
\texttt{\small{roberto.passerone@unitn.it}}
}

\date{\small{\textsuperscript{1}Department of Information Engineering and Computer Science \\ University of Trento \\ via Sommarive 9, 38123 Povo, Trento, Italy}
}

\maketitle

\begin{abstract}
\noindent
This paper introduces cuVegas, a CUDA-based implementation of the Vegas Enhanced Algorithm (VEGAS+), optimized for multi-dimensional integration in GPU environments. The VEGAS+ algorithm is an advanced form of Monte Carlo integration, recognized for its adaptability and effectiveness in handling complex, high-dimensional integrands. It employs a combination of variance reduction techniques, namely adaptive importance sampling and a variant of adaptive stratified sampling, that make it particularly adept at managing integrands with multiple peaks or those aligned with the diagonals of the integration volume. Being a Monte Carlo integration method, the task is well suited for parallelization and for GPU execution. Our implementation, cuVegas, aims to harness the inherent parallelism of GPUs, addressing the challenge of workload distribution that often hampers efficiency in standard implementations. We present a comprehensive analysis comparing cuVegas with existing CPU and GPU implementations, demonstrating significant performance improvements, from two to three orders of magnitude on CPUs, and from a factor of two on GPUs over the best existing implementation. We also demonstrate the speedup for integrands for which VEGAS+ was designed, with multiple peaks or other significant structures aligned with diagonals of the integration volume.

\vspace*{10pt}
\end{abstract}

\setcounter{footnote}{0}
\renewcommand{\thefootnote}{\alph{footnote}}

\vspace*{1pt}\textlineskip    

\section{Introduction}\label{intro}
Numerical integration is a fundamental computational technique 
widely used to approximate the definite integral of functions,
especially when analytical solutions are difficult or impossible to obtain. 
It plays a vital role in computational science, 
spanning a wide array of fields from physics to finance,
making it a critical component in scientific computing and real-world applications~\cite{davis2007methods}.
In computational physics, numerical integration is employed to simulate complex physical systems, 
enabling the precise calculation of volumes~\cite{KalosWhitlock2008}. 
In finance, numerical integration plays a vital role in the pricing of derivative securities 
and risk assessment through techniques such as Monte Carlo simulations~\cite{Glasserman_2010}. 
Moreover, Bayesian parameter estimation extensively utilizes numerical integration to compute posterior distributions, 
which is essential for making informed probabilistic inferences~\cite{Gelman2013}.

There are several methods to compute integrals for numerical integration,
each suited to specific types of problems and computational constraints. 
Traditional methods include the Trapezoidal Rule and Simpson's Rule, 
which are often employed for problems with smooth integrands and lower dimensions, 
partitioning the integration domain into smaller segments 
and approximating the integral by fitting simple geometric shapes~\cite{Burden2015}. 
Another classical approach is the Gaussian Quadrature, 
which selects optimal points and weights for integration to achieve high accuracy with fewer function evaluations~\cite{10.1145/108556.108580, davis2007methods}.
These traditional methods are effective for low-dimensional problems, 
but often become computationally infeasible for high-dimensional integrals or integrals with complex boundaries.

Among various techniques, Monte Carlo methods
stand out for their effectiveness in dealing with multidimensional integration
problems, or complex domains~\cite{rubinstein2016simulation}.
These methods estimate integrals by averaging the values of the integrand function 
at randomly selected points within the integration domain.
Notably, the precision of Monte Carlo methods improves with the increase in the
number of function evaluations, which, however, can lead to slow convergence
in complex, high-dimensional cases~\cite{Lepage_2021, Liu_2001}.
A key feature of Monte Carlo integration is its general applicability; it
operates effectively without stringent requirements on the integrand properties.
The method does not mandate the integrand to be either analytic or continuous
and also provides dependable uncertainty estimations.
This flexibility proves advantageous in multidimensional integration,
especially when critical aspects of the integrand are confined to small areas
of the integration space, highlighting the need for adaptable integration
strategies.

The VEGAS algorithm represents a notable advancement in adaptive
multidimensional integration~\cite{PETERLEPAGE1978192}.
Originating from the domain of particle physics for Monte Carlo simulation 
and Feynman diagram evaluation~\cite{Kersevan_Richter-Was_2013, Alwall_et_al_2014, Aoyama_et_al_2012}, 
VEGAS has progressively broadened its applicability. Its use extends to chemical physics for path integrals 
and virial coefficient calculations~\cite{Garberoglio_Harvey_2011}, 
and to finance for complex option pricing models~\cite{Campolieti_Makarov_2007}. 
In astrophysics, VEGAS aids in high-redshift supernova data analysis 
and galactic dynamics studies~\cite{Serra_Heavens_Melchiorri_2007, Sanders_2014, Gultekin_et_al_2009}. 
Furthermore, it serves computational neuroscience in neural network modeling~\cite{Atay_Hutt_2006}, 
and atomic physics in quantum entanglement exploration~\cite{Dehesa_et_al_2012}. 
This extensive applicability underscores the algorithm capability 
to address complex computational problems across diverse scientific domains.

The adaptability of the VEGAS algorithm is very effective for functions with pronounced peaks, 
but functions with multiple peaks or peaks aligned with the integration volume diagonals require further enhancements. 
The VEGAS Enhanced Algorithm (VEGAS+)~\cite{Lepage_2021} integrates adaptive importance sampling 
with a variant of adaptive stratified sampling, augmenting its efficiency for complex functions. 
Despite its effectiveness, sequential execution can result in extended computation times, 
a limitation overcome through parallelization on GPUs.

The advent of powerful computational hardware, particularly Graphics Processing Units (GPUs), 
has revolutionized the efficiency and scalability of Monte Carlo integration algorithms~\cite{10.5555/1964878}. 
GPUs are well-suited for parallel computing tasks due to their massive parallelism capabilities, 
making them ideal for executing Monte Carlo simulations that involve large-scale sampling~\cite{Owens2008}.
By implementing Monte Carlo integration on GPUs, researchers can achieve significant speed-ups 
and handle computationally intensive tasks more effectively, which opens up new possibilities 
in real-time data analysis and simulation~\cite{Kanzaki_2011}. 

This paper introduces cuVegas, a CUDA-based implementation of the VEGAS+ algorithm 
specifically designed for multi-dimensional integration on GPU architectures. 
The cuVegas implementation takes full advantage of the parallel processing capabilities inherent in GPUs, 
significantly accelerating the computation process while addressing the challenges of uneven workload distribution 
typically encountered in traditional implementations. 
This is achieved through a novel load-balancing approach 
that ensures an even distribution of computational tasks across GPU threads, 
thereby maximizing efficiency and minimizing idle time.

In addition to optimizing workload distribution, cuVegas features an on-GPU map update mechanism 
that enhances the adaptive stratified sampling process, a key aspect of the VEGAS+ algorithm. 
Furthermore, our implementation supports multi-GPU environments, 
making it adaptable to modern computational setups where multi-GPU systems are becoming increasingly common. 
By leveraging multiple GPUs, cuVegas can exploit additional parallelism, further boosting performance and scalability.
To enhance usability and accessibility, we have also developed a Python binding for cuVegas. 
This binding simplifies the integration of our implementation into existing workflows,
allowing to easily incorporate GPU-accelerated multi-dimensional integration into their Python-based applications.

Our evaluation of cuVegas is comprehensive, extending beyond standard benchmarks to include relevant real-world applications. 
We demonstrate its efficacy through extensive numerical experiments, 
including the simulation of Feynman path integrals in quantum physics and the pricing of Asian options in financial mathematics. 
These applications showcase the practical utility of our implementation in solving complex problems across diverse domains.
In terms of performance, cuVegas achieves a computational speedup of up to a factor of 20 compared to previous GPU-based methods. 
Our benchmarks and tests show that cuVegas outperforms existing VEGAS Enhanced Algorithm implementations 
by achieving the same level of accuracy in less time.
The results demonstrate that cuVegas is a highly efficient tool for numerical integration, 
making it an essential resource for complex multi-dimensional integration tasks.

The subsequent sections of this paper are organized as follows: 
Section~\ref{background} elucidates the VEGAS algorithm, 
its foundational methodology, and associated challenges; 
Section~\ref{implement} elaborates on our CUDA-focused implementation strategy; 
Section~\ref{test} showcases the efficacy of our implementation 
through test scenarios involving both test integrands and real-world applications. 
We conclude the paper with Section~\ref{concl}.

\section{Background and Methodology} \label{background}
In this section, we discuss the foundational aspects and methodologies of the VEGAS+ algorithm and related work, followed by the challenges faced in parallelizing the algorithm for GPU implementations.

\subsection{The VEGAS Enhanced Algorithm}
VEGAS+ is an adaptive and iterative algorithm which refines its approximation based on
prior iterations.
Specifically, the algorithm partitions each axis into grids, thereby dividing
the integration space into hypercubes.
Monte Carlo integration is then performed within each hypercube, and the
related variance of the integrals
is used to adapt the grid for
subsequent iterations.
This adaptability is achieved through two primary variance reduction
techniques: \emph{adaptive importance sampling} and \emph{adaptive stratified
sampling}.

\subsubsection{Adaptive Importance Sampling}
The core idea behind adaptive importance sampling is to focus computational
resources on those regions of the domain that contribute most significantly to
the integral.
The algorithm achieves this by transforming the original integral using a
Jacobian matrix, effectively ``flattening'' the integrand.
For a one-dimensional integral of the form:
\begin{equation}
    \int_{a}^{b} f(x) \,dx,
\end{equation}
the transformation yields:
\begin{equation}
    \int_{0}^{1} J(y) f(x(y)) \,dy,
\end{equation}
where \( J(y) \) is the Jacobian of the transformation
realized as a step function based on the VEGAS map.
The map divides the $x$-axis into $N_g$ intervals that map
\( [0, 1] \) in the $y$-space to \( [a, b] \) in the $x$-space.
Uniform intervals of width $1/N_g$ map to intervals of width $\Delta x_i$.
The Jacobian is then written as follows:
\begin{equation}
    J(y) = N_g \Delta x_{i(y)},
\end{equation}
where $y$ is the transformed variable in \( [0, 1] \), and $i(y)$ is the integer part of $yN_g$.
The map is refined through the iterations varying the interval sizes,
controlled by a damping parameter $\alpha$, so that the variance of the
estimate is minimized by concentrating samples in the peaked regions of the
integrand.
The Monte Carlo estimate \( I_{MC} \) is then given by:
\begin{equation}
    I_{MC} = \frac{1}{N_{ev}}\sum_{y} J(y) f(x(y)),
\end{equation}
where \( N_{ev} \) is the number of uniformly sampled points in the interval
\( [0, 1] \).

\subsubsection{Adaptive Stratified Sampling}
The original stratification of the VEGAS algorithm considers a uniform
distribution of the integrand evaluations in each hypercube.
In contrast, adaptive stratified sampling aims to distribute the number of
function evaluations across the hypercubes, in a manner that minimizes the
overall variance of the integral.
The integral estimate \( I_{MC} \) and variance \( \sigma^2_{MC} \) are
computed as:
\begin{align}
    I_{MC} &\approx \sum_{h} \Delta I_h, \\
    \sigma^2_{MC} &\approx \sum_{h}  \frac{\sigma^2_h(Jf)}{n_h},
\end{align}
where \( \Delta I_h \) is the contribution of each hypercube, and
$\sigma^2_h(Jf)$ is the variance of the product of the Jacobian with the
integrand samples for each hypercube.
The optimal number of evaluations per hypercube \( n_h \) is proportional to
\( \sigma_h (Jf) \), subject to the constraint:
\begin{equation}
    N_{eval} = \sum_{h} n_h,
\end{equation}
where \( N_{eval} \) is the total number of evaluations.
A damping parameter \( \beta \) is introduced to mitigate fluctuations,
enhancing the algorithm performance for integrands with multiple peaks.

As a result, the number of function evaluations varies across hypercubes,
leading to a more efficient integral evaluation.
This is especially true for integrands with complex peaked structures or
diagonal peaks, where the evaluation points are more concentrated in the parts
of the domain that correspond to the to peaks.

\subsubsection{Estimation Aggregation}
Finally, the integral estimate is refined by aggregating the estimates from
various iterations through a weighted average, taking into account the
variance of these estimates.
The overall integral estimate and variance are computed as:
\begin{align}
    I &= \frac{\sum_{i} \frac{I_i}{\sigma^2_i}}{\sum_{i} \frac{1}{\sigma^2_i}}, \\
    \sigma^2 &= \frac{1}{\sum_{i} \frac{1}{\sigma^2_i}}.
\end{align}

\subsection{Related Work}
The VEGAS algorithm is implemented across multiple languages and frameworks,
including Python packages, C++ libraries such as Cuba~\cite{HAHN200578} and
the GNU Scientific Library GSL~\cite{10.5555/1538674}, and the R Cubature package~\cite{cubature}.
Despite its efficiency for certain applications, the algorithm sequential
execution often leads to lengthy computation times, a limitation addressed by
employing parallelization, especially GPUs.
Kanzaki initially suggested a CUDA GPU implementation, gVEGAS~\cite{Kanzaki_2011},
offering a 50x speedup with respect to a comparable C implementation on CPU.
The program processes each hypercube in a single thread, which could cause work imbalance.
Moreover, the update of the importance sampling map is performed on the CPU,
which can be inefficient.
Sakiotis et al.\ later introduced
m-CUBES~\cite{DBLP:journals/corr/abs-2202-01753}, an optimized GPU version
that achieves uniform workload distribution across the parallel processors by
assigning batches of cubes to each parallel thread.
This technique is effective for VEGAS, since to each hypercube is assigned an
equal number of integrand evaluations, unlike VEGAS+, which employs adaptive
stratified sampling.

For the VEGAS+ algorithm, the original program by Lepage is
Vegas~\cite{peter_lepage_2023_8175999}, a Cython implementation that can
utilize multiple CPUs.
GPU adaptations include VegasFlow~\cite{Carrazza:2020rdn}, a Python package
based on the TensorFlow library~\cite{vegasflow_package}, and
TorchQuad~\cite{Gómez2021}, which utilizes PyTorch and it is also available as
a Python package ~\cite{torchquad_package}.
Both packages offer also other integration algorithms.
These Python libraries use CUDA for computational tasks by leveraging their
respective frameworks, but may introduce overhead compared to a native CUDA
implementation, which offers potential for enhanced optimization and more
effective hardware utilization, by being more tailored to the task.

The performance of cuVegas will be benchmarked against
Vegas~\cite{peter_lepage_2023_8175999}, VegasFlow~\cite{vegasflow_package} and
TorchQuad~\cite{torchquad_package}, to assess its efficiency and optimization
relative to existing solutions.
Moreover, we also provide a comparison with m-CUBES~\cite{mcubes} to
investigate the advantages of considering VEGAS+.

\subsection{Parallelization challenges}
Implementing the VEGAS+ algorithm on CUDA introduces significant
challenges, chiefly in optimizing integrand evaluation to fully leverage GPU
resources.
As the computational demand of the integrand escalates, its evaluation
predominates the computation time.
Achieving full Streaming Multiprocessor (SM) occupancy during this process is
critical, necessitating a workload balance to utilize the parallel processors
efficiently.

A naive parallelization strategy consists in assigning hypercubes to GPU
threads, facilitating the accumulation of weights for the importance sampling
map within each hypercube.
However, this method encounters work imbalance due to the adaptive
stratification characteristic of VEGAS+.
Such imbalance is particularly pronounced for integrands with sharp peaks, as
the sampling map becomes highly irregular, leading to thread divergence.

In addition, this approach requires a thread for each hypercube,
with the number of threads related to the number of stratifications in each dimension.
Therefore, the required grid-size could be prohibitive,
and the algorithm parameters could be restricted by memory constraints.

The parallelization challenges of the VEGAS+ algorithm can be summarized as follows:
\begin{itemize}
\item \textbf{Integrand evaluation}\\
The computational intensity of the integrand varies, but typically, it is the
    most demanding aspect of the algorithm. In this stage, it is crucial to
    exploit the parallel threads, by designing a balanced partitioning scheme
    that allows an efficient workload distribution avoiding thread divergence.
\item \textbf{Random number generation}\\
Random number generation (RNG) is essential for sampling points in the domain
    for function evaluation. RNG requires to be both random and efficient,
    necessitating a balance between the number of RNGs and memory constraints.
\item \textbf{Memory access patterns}\\
The adaptive algorithm requires access to grids and maps of previous
    iterations when sampling the points. To improve performance and limit
    latency, correct access patterns for threads performing sampling are
    crucial.
\item \textbf{Results accumulation}\\
The algorithm adaptive behavior requires computing weights by accumulating
    results from multiple integrand evaluations based on their relation to
    importance sampling and stratification maps.
    This needs an accumulation strategy for the results of each evaluation into the maps.
    Furthermore, the final estimate is obtained by summing all intermediate results within each
    hypercube, which could benefit from parallelization.
\item \textbf{Memory transfers and map update}\\
Limiting memory transfer operations between the host and GPU device is crucial for
    performance. Enabling GPU computation for non-fully parallel
    operations, such as map updates involving sequential steps, is essential.
    With data already residing on GPU, sub-optimal GPU computation might still be faster than memory transfers to CPU.
\end{itemize}

\section{CUDA Implementation of the VEGAS Enhanced Algorithm} \label{implement}
This section provides an in-depth exploration of our CUDA-based implementation
of the VEGAS Enhanced Algorithm.
Our implementation is based on CIGAR~\cite{cigar} a single-threaded C++
implementation of VEGAS/VEGAS+.
The source code for the proposed CUDA implementation is publicly accessible~\cite{vegas}.
We discuss the architecture, optimization strategies, and performance
considerations that contribute to the robustness and efficiency of our
implementation.

\subsection{Parallelization strategy}
The naive parallelization method, while effective in certain scenarios such as
non-peaked integrands and when the number of hypercubes aligns with GPU
grid-size constraints, has limitations.
This is mainly due to the adaptive stratification of VEGAS+,
which can introduce work imbalance among the hypercubes,
leading to thread divergence and consequently slow execution times for peaked functions.
A more robust approach, aligning with the adaptive nature of VEGAS+, involves
parallelizing \emph{batches} of integrand evaluations.
By employing batches of function evaluations,
the parallelization scheme becomes more robust,
eliminating work imbalance by ensuring each thread
executes the same number of random point samplings and integrand evaluations.

Moreover, batching enhances flexibility by permitting different batch sizes,
allowing adjustments in the number of threads and meeting memory constraints.
This method utilizes a predefined grid size for random number generation and
simultaneous function evaluations, thus reducing the number of RNGs and
limiting related memory usage, while facilitating the parallel execution of
function evaluations.
However, this approach requires mapping these evaluations to their
corresponding hypercubes based on the stratification map, which is a non
fully-parallel operation.
Moreover, this approach introduces the need to accumulate the results of the
function evaluations to the related hypercube.
The accumulation of weights for importance sampling and stratification maps
necessitates atomic operations to ensure accuracy and consistency.

\begin{figure}[ht!]
    \centering
    \includegraphics[trim={0 8cm 0 5.5cm},clip,width=0.9\linewidth]{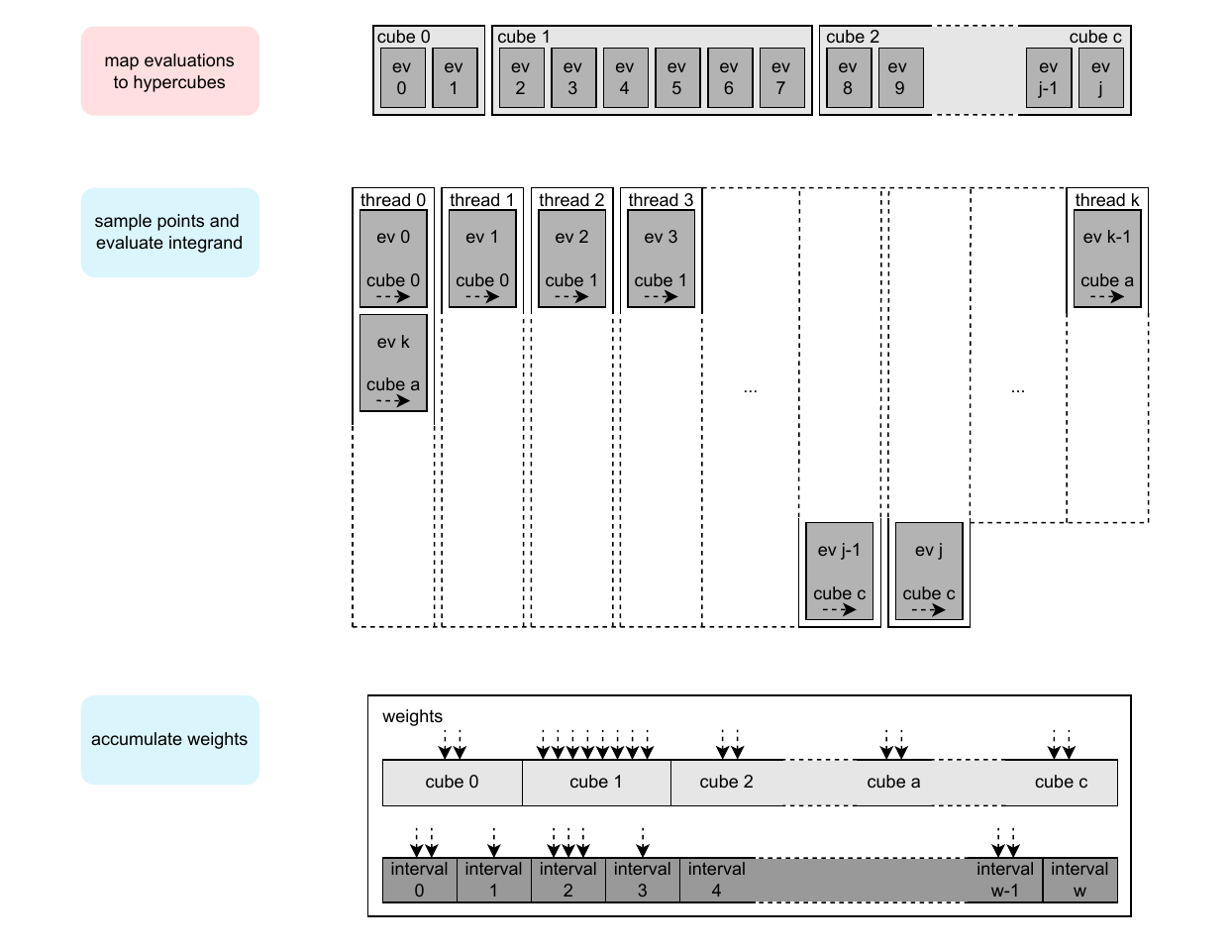}
    \caption{Parallelization diagram of the program in a single GPU setting.} \label{fig:schema_single}
\end{figure}

\begin{figure}[ht!]
    \centering
    \includegraphics[trim={0 5cm 0 0.5cm},clip,width=0.9\linewidth]{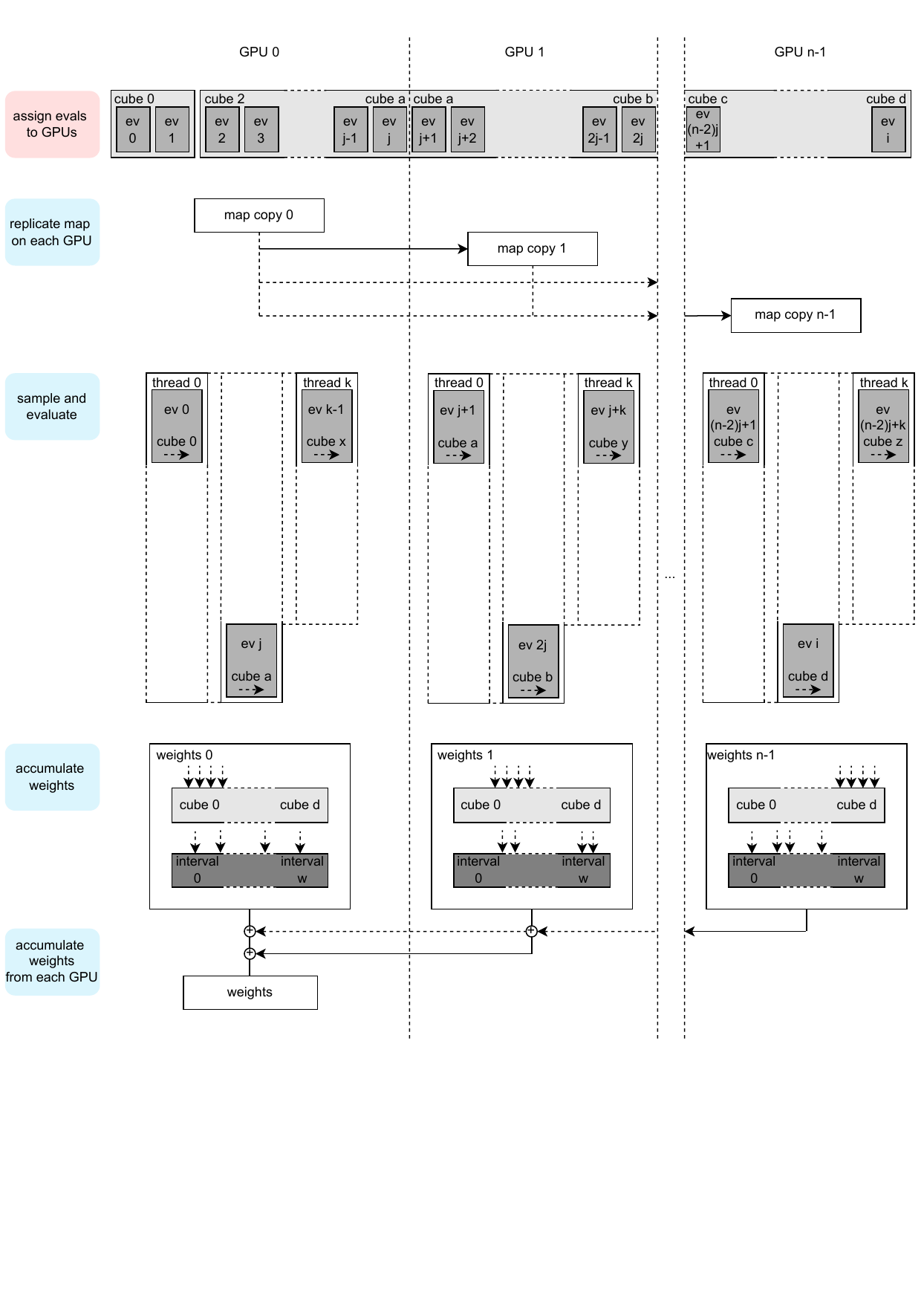}
    \caption{Parallelization diagram of the program in a multi GPU setting.} \label{fig:schema_multi}
\end{figure}

The parallelization strategy is further represented in Figure~\ref{fig:schema_single} for single GPU execution and in
Figure~\ref{fig:schema_multi} for multi-GPU scenarios.
In detail, the schemas begin with
the mapping of evaluations to hypercubes, which,
starting from the number of function evaluations for each hypercube,
allows each of the function evaluations to have a reference to its hypercube.
In case of multi-GPU execution, the function evaluations are distributed evenly
across the available devices and the auxiliary data structures 
are also replicated in parallel,
starting from the first GPU to all GPUs.
After the mapping, the $j$ evaluations are processed by $k$ logical threads
corresponding to the batch size, so that each thread computes about the same number
of random integrand points ($\pm1$ when $j$ is not multiple of $k$).
After the computation of the weights, they need to be accumulated in the required maps,
according to the related hypercubes and interval indexes.
Those values, residing in different threads, are accumulated via atomic instructions.
For multi-GPU execution, the maps are accumulated as a parallel reduction,
so that the actual total map resides on the first GPU.

Although this strategy introduces overhead due to the mapping of evaluations
to hypercubes and the accumulation of hypercube weights, it significantly
enhances resource utilization.
This results in more consistent performance and scalability, particularly for
handling highly peaked integrands, which are a focus of the VEGAS+ algorithm.

The process of updating the importance sampling map is inherently sequential
and cannot be fully parallelized.
Nonetheless, performing this update on the GPU offers a performance advantage
over CPU execution.
This advantage stems from eliminating the need for memory transfers between
the device and the host, thereby enhancing overall efficiency.

\subsection{Algorithm Overview and Pseudocode}

\begin{algorithm}
\caption{Optimized Algorithm Pseudocode} \label{alg:program}
    \bboxit{myyellow}{1.05}
    \textit{init}( )\tcp*{initialize memory and variables}
    \textit{states $\gets$ setupRNG}( )\tcp*{setup RNG states \textbf{parallel}}
    \ForAll{\textit{it} \textbf{in} \textit{iterations}}{
        \bboxit{mygreen}{1.05}
        \lForEach(\tcp*[f]{reset weights \textbf{parallel}}){\textit{w} \textbf{in} \textit{mapWeights}}{\textit{w $\gets 0$}}
        \lForEach(\tcp*[f]{reset weights \textbf{parallel}}){\textit{w} \textbf{in} \textit{cubWeights}}{\textit{w $\gets 0$}}
        \bboxit{myred}{0.05}
        \textit{evals $\gets$ mapEvalsToCubes}(\textit{nh})\tcp*{map evaluation to related cube \textbf{parallelized}}
        \bboxit{myblue}{6.15}
        \ForAll(\tcp*[f]{fill weights \textbf{parallel}}){\textit{state} \textbf{in} \textit{states}}{
            \textit{runs $\gets$ getRuns}(\textit{evals})\tcp*{get runs for this instance}
            \textit{mapWeightsTmp, cubWeightsTmp $\gets$ vegasFill}(\textit{runs, state, map, mapWeights, cubWeights})\tcp*{fill}
            \textit{mapWeights, cubWeights $\gets$ mapWeights, cubWeights + mapWeightsTmp, cubWeightsTmp}\tcp*{accumulate}
        }
        \bboxit{mygreen}{7.15}
        \textit{nh $\gets$ updateEvalsPerCube}(\textit{cubWeights, beta})\tcp*{compute stratification map \textbf{parallel}}
        \textit{mapWeights $\gets$ adjustWeights}(\textit{mapWeights, alpha})\tcp*{smooth map weights \textbf{parallel}}
        \textit{map $\gets$ updateMap}(\textit{map, mapWeights})\tcp*{update importance sampling map \textbf{parallelized}}
        \If{\textit{it $>$ skip}} {
            \textit{res, sig $\gets$ computeResults}(\textit{cubWeights})\tcp*{accumulate results from cubes \textbf{parallel}}
            \textit{results}[\textit{it}]\textit{, sigmas}[\textit{it}]\textit{ $\gets$ res, sig}\tcp*{store iteration result}
            \textit{result, sigma2 = combineIterationsResults}(\textit{results, sigmas})\tcp*{compute result}
        }
    }
    \bboxit{myorange}{0.05}
    \textit{clear}( )\tcp*{release memory}
\Return \textit{result, sigma2}\;
\end{algorithm}

\begin{algorithm}
\caption{VegasFill} \label{alg:vegasfill}
    \bboxit{myblue}{8.15}
    \ForAll{\textit{run} \textbf{in} \textit{runs}}{
        \textit{cub $\gets$ getCube}(\textit{run})\tcp*{get cube for the run}
        \textit{x, id $\gets$ getPoint}(\textit{state, cub})\tcp*{sample point in the cube}
        \textit{j $\gets$ getJac}(\textit{id})\tcp*{compute jacobian}
        \textit{f $\gets$ integrand}(\textit{x})\tcp*{evaluate integrand}
        \textit{mapWeights}[\textit{id}]\textit{ $\gets$ mapWeights}[\textit{id}]\textit{ + GetMapWeights}(\textit{j, f})\tcp*{accumulate map weights}
        \textit{cubWeights}[\textit{cub}]\textit{ $\gets$ cubWeights}[\textit{cub}]\textit{ + GetCubWeights}(\textit{j, f})\tcp*{accumulate cube weights}
    }
\Return \textit{mapWeights, cubWeights}\;
\end{algorithm}

The program, outlined in Algorithm~\ref{alg:program},
starts with the
initialization of algorithmic parameters and memory allocation (highlighted in yellow).
The random number generators (RNGs) are set up using the
cuRAND library \verb|XORWOW| algorithm.
The main loop iterates through the VEGAS algorithm, refining the integral
estimation and adapting the grid for enhanced accuracy.
The weights of the contribution of each integrand sample,
used for updating the map and stratification, are reset (highlighted in green), and every needed
integrand function evaluation is mapped to its respective hypercube (highlighted in red), to allow
a parallelization with respect to every function evaluation for the successive
\emph{vegasFill} routine (highlighted in blue), outlined in Algorithm~\ref{alg:vegasfill}.
For every evaluation in each hypercube, the program generates a random point
in the transformed space for which the value of the integrand function is
evaluated.
Then the Jacobian
and weights of that evaluation are computed
and accumulated to update the map and stratification parameters.

During the update process (highlighted in green), the accumulated weights are utilized
in the \emph{updateEvalsPerCube} phase to determine the effective number of function
evaluations per hypercube.
Following this, the map is updated to the new map by computing new parameters.
This involves normalization, smoothing and aggregation of the weights with the
counts collected in earlier phases.

Simultaneously to the map update, if the
iteration index is above a threshold, which is set for the convergence of the
map and stratification parameters,
the program computes the new estimate of the integral.
The results within each hypercube are accumulated in the \emph{computeResults} routine,
to obtain the
estimate of the integral for the current iteration, as well as its variance.
Subsequently, the results from the previous iterations are combined to obtain
a weighted approximation for the integral outcome.
After the completion of all iterations, the memory is deallocated (highlighted in orange), 
and the ultimate integral estimate becomes the program output.

\subsection{Implementation details and optimizations}

To overcome the parallelization challenges and devise an effective strategy,
we must first assess the running time of the various sections of the algorithm
and identify areas that require optimization.
With this objective in mind, we meticulously examine the implementation
details of the algorithm, focusing on techniques to enhance performance and
scalability.
By understanding the details of the algorithm execution on GPU architectures,
we propose optimizations that mitigate bottlenecks and improve overall
efficiency.

\subsubsection{Program time analysis}

Table~\ref{tab:breakdown} reports the running time of the different sections
of the algorithm.
We considered both a computationally ``easy'' function (Roos\&Arnold)
and an ``intensive'' and peaked function (Ridge), defined in Table~\ref{tab:integrand},
by examining also the scalability in the number of function evaluations.
The majority of the time is
spent on the \emph{vegasFill} routine, where the integrand is evaluated and
the results are accumulated. As expected, increasing the number of integrand
evaluations translates to a higher
total time fraction spent on the filling
part, even for computationally ``easy'' integrands, while it
becomes evident for ``intensive'' integrands.
Therefore, it is important to optimize the filling section by allowing a full
resource optimization of the integrand evaluations.
The other sections benefit also from parallelization but the improvement is
minor.

\begin{table}[ht!]
    \caption{Breakdown table of the program running time with multiple number
    of evaluations of integrand functions of Table~\ref{tab:integrand}. The
    parameters are set to the ``def'' configuration with reference to
    Table~\ref{tab:config}. The color scheme matches the sections of
    Algorithm~\ref{alg:program}. The reported time ignores the CUDA context
    initialization.}
    \label{tab:breakdown}
    \begin{tabular}{|c||c|c|c|c||c|c|c|c|}
        \hline
        \textit{integrands} & \multicolumn{4}{c||}{Roos\&Arnold} & \multicolumn{4}{c|}{Ridge} \\
        \cline{1-9}
        integrand evaluations & $10^7$ & $10^8$ & $10^9$ & $10^{10}$ & $10^7$ & $10^8$ & $10^9$ & $10^{10}$\\
        \hline
        \cellcolor{myyellow!25}
        init [\%] & 18.2 & 10.6 & 4.4 & 2.3 & 4.1 & 0.9 & 0.3 & 0.2\\
        \cellcolor{myred!25}
        map [\%] & 10.4 & 5.2 & 3.3 & 3.8 & 2.2 & 0.5 & 0.3 & 0.3\\
        \cellcolor{myblue!25}
        fill [\%] & \textbf{35.8} & \textbf{70.3} & \textbf{88.0} & \textbf{91.7} & \textbf{89.2} & \textbf{97.7} & \textbf{99.1} & \textbf{99.3}\\
        \cellcolor{mygreen!25}
        update [\%] & 33.0 & 12.0 & 3.8 & 2.1 & 3.9 & 0.7 & 0.2 & 0.1\\
        \cellcolor{myorange!25}
        clear [\%] & 2.6 & 1.9 & 0.5 & 0.1 & 0.6 & 0.2 & 0.1 & 0.1\\
        \hline
        total time [ms] & 29.0 & 91.5 & 570.8 & 4697.2 & 149.4 & 1123.1 & 10888.8 & 108170.2\\
        \hline
    \end{tabular}
\end{table}

\subsubsection{Optimization of Random Number Generation}
The \verb|vegasFill| procedure, invoked through a kernel launch, is the most
computationally demanding part of our algorithm, requiring the highest number
of work units.
This procedure begins by randomly sampling points in the transformed space,
utilizing the cuRAND library \verb|XORWOW| algorithm for random number
generation.
Each thread is assigned its own \verb|XORWOW| state during the \emph{setupRNG}
phase, initialized with a uniform seed but varying sequence numbers.

Given that state initialization is resource-intensive, both in terms of time
and memory, it is impractical to allocate a unique state for each evaluation
point.
To address this, we introduce a \verb|batch_size| parameter.
This parameter specifies the maximum number of threads that the
\verb|vegasFill| kernel can employ, thereby determining the number of
\verb|XORWOW| states that need to be initialized.
Consequently, each thread is tasked with generating
\(\frac{n_{\text{eval}}}{\text{batch\_size}}\) random points.

With each \verb|curandState| consuming 48 bytes, the upper limit for
\verb|batch_size| is approximately \(10^9\), subject to the available RAM on
the GPU.
Additionally, to maximize the utilization of the GPU Streaming Multiprocessors
(SMs), the \verb|batch_size| must be sufficiently large.
It is worth noting that the initialization of RNG states is a one-time
operation, thereby incurring a fixed computational overhead that is amortized
over the entire runtime of the algorithm.

\subsubsection{Accumulation and Data Reduction Techniques}
The \verb|vegasFill| procedure leverages \verb|AtomicAdd| instructions to
aggregate weights for both the map and stratification, corresponding to each
evaluation specific map interval and hypercube.
Reduction operations are also needed for the computation of the integral
estimate and variance by accumulating the hypercubes results, and for the map
update.
These accumulations are further optimized through the use of the CUB software
library~\cite{cub}, which offers device-wide specialized functions for array
summation, including reduction and prefix-scan operations.

\subsubsection{Mitigating Thread Divergence in Hypercube Mapping}
Our algorithm parallelization strategy necessitates mapping each integrand
function evaluation to its designated hypercube.
While GPU-compatible, this operation is prone to thread divergence due to the
variable number of evaluations allocated to each hypercube.
To counteract this, we implement a conditional approach: the mapping operation
is executed on the CPU if the maximum fraction of total evaluations for a
single hypercube surpasses a predetermined threshold.
Otherwise, it remains on the GPU.
For CPU-based mapping, we employ OpenMP parallelization when the evaluation
count exceeds a specific value, thereby optimizing multi-threading overhead.

\subsubsection{Stream-Based Parallelization for Enhanced Performance}
The majority of the computational workload within the main loop is executed on
the GPU.
To further enhance performance, we employ CUDA streams to parallelize memory
operations and computations.
Specifically, map updates, which include weight normalization and smoothing,
are processed on a high-priority stream.
This is separate from the stream responsible for integral estimation, variance
calculation, and stratification updates.

\subsection{Multi-GPU approach} \label{sec:multi}
The implementation enables the execution across multiple GPUs for the \verb|vegasFill| kernel,
responsible for integrand calls and typically consuming the majority of the execution time.
Other computations, such as parameter updates and result computation, are handled on a single device.
The RNG is initialized by assigning a distinct offset in the random sequence to each GPU device,
ensuring an adequate number of random number generations without sequence overlap.

The communication between GPUs is obtained with the \verb|cudaDeviceEnablePeerAccess|
function of the CUDA runtime API, which leverages the \emph{NVLink} interconnection if possible
(available in the SXM design used for testing), to access the peer memory without involving the CPU.
Data sharing is achieved through peer-to-peer communication (P2P) for the purpose of merging the fill results.
Prior to the kernel call, global memory data structures are replicated to each GPU device,
enhancing performance using a tree-like broadcast strategy, which involves
\( \lceil \log_2{n} \rceil \) \verb|deviceToDevice| data copy steps for \( n \) devices.
The same strategy is adopted for merging results which requires summing the values into a single array.
These data replications and accumulation from multiple GPUs are reported in the parallelization schema
in Figure~\ref{fig:schema_multi}.

Utilizing multiple GPUs can be advantageous, particularly with highly computation-intensive integrands,
where the data sharing overhead is minimal,
and the overall speedup approaches the number of participating devices.

\subsection{Performance limitations}
The \verb|vegasFill| kernel accounts for the majority of the program overall
execution time.
As outlined in the pseudocode~\ref{alg:vegasfill}, it is responsible for the
random sampling of the domain points, integrand function evaluation, and
weights accumulation with atomic functions.
Its computation intensity depends on the integrand device function which is
executed in the \verb|vegasFill| kernel, and typically results in a
double-precision floating-point computationally-bound execution.
Moreover, the number of required registers, which is also integrand function
dependent, limits the occupancy, and small block sizes typically show better
overall performance.

Each function evaluation requires the access to the importance sampling map,
to locations based on the randomly sampled domain point in each thread.
This almost random global memory access introduces latency.
A possible approach to mitigate this could be sampling the points beforehand
and then order the evaluations to improve coalescing for global memory access
to reduce the load on the memory controller.

\subsection{Python binding}
In order to enhance the program usability and efficiency, we developed a
CUDA PyTorch extension using pybind11~\cite{pybind11}, a lightweight
header-only library that acts as a connection between C++ and Python.
This extension enables seamless integration of the existing CUDA C program
with the PyTorch framework.

The binding process involves overseeing minimal constant overhead, measurable
from 50 to 100~ms, primarily attributed to the just-in-time (JIT) compilation
of the integrand function to PTX code, facilitated by the Numba
package~\cite{10.1145/2833157.2833162}, wherein the integrand is defined as a
Numba CUDA device function in Python code.
This compilation ensures the integrand function can be linked at runtime to
the previously compiled segments of the program.
This approach incurs minimal overhead, while enabling a smoother and more
straightforward implementation of the algorithm within the Python context.

\section{Performance Analysis} \label{test}
In this section, we delve into a comprehensive performance analysis of
cuVegas.
We benchmark cuVegas against existing implementations, including the original
CPU-based Vegas, TorchQuad, and VegasFlow.
Our analysis covers multiple dimensions, such as computational speed, memory
usage, and the impact of various optimization techniques.
The goal is to provide a holistic view of the performance characteristics and
advantages of cuVegas.

\subsection{Methodology} \label{methodology}

\begin{table}[ht!]
  \caption{Configuration Parameters}
  \label{tab:config}
  \begin{tabular}{c|c|c|c}
    \toprule
    Parameter & Configuration 1 (def) & Configuration 2 (vf) & Configuration 3 (tq) \\
    \midrule
    max\_it & 20 & 20 & 20 \\
    skip & 0 & 0 & 0 \\
    max\_batch\_size & 1048576 & 1048576 & 1048576 \\
    n\_intervals & 1024 & 50 & computed on n\_eval \\
    alpha & 0.5 & 1.5 & 0.5 \\
    beta & 0.75 & 0.75 & 0.75 \\
    \bottomrule
  \end{tabular}
\end{table}

\begin{table}[ht!]
  \caption{Test Integrands in This Study}
  \label{tab:integrand}
  \begin{tabular}{llcl}
    \toprule
    \# & Integrand & Dimensions (d) & Function\\
    \midrule
    (1\label{sin_exponential}) & \emph{Sine Exponential} & 2D & $f(x) = \sin(x_1) + \exp(x_2)$\\[4ex]
    (2\label{linear}) & \emph{Linear} & 10D & $f(x) = \sum_{i=1}^{d} x_i$\\[4ex]
    (3\label{cosine}) & \emph{Cosine} & 10D & $f(x) = \prod_{i=1}^{d} \cos(x_i)$\\[4ex]
    (4\label{exponential}) & \emph{Exponential} & 10D & $f(x) = \exp(\sum_{i=1}^d x_i^2)$\\[4ex]
    (5\label{roos}) & \emph{Roos \& Arnold} & 10D & $f(x) = \prod_{i=1}^{d}\left|4 x_i - 2 \right|$\\[4ex]
    (6\label{Morokoff}) & \emph{Morokoff \& Caflisch} & 8D & $f(x) = (1 + 1/d)^d\prod_{i=1}^d x_i^{1/d}$\\[4ex]
    (7\label{gaussian}) & \emph{Gaussian} & 4D & $f(x) = \frac{1}{(2 \pi \sigma^2)^{d / 2}}\exp \left( -\frac{\sum_{i=1}^d (x_i - \mu)^2}{2 \sigma^2} \right); \mu = 0.5, \sigma = 0.01$\\[4ex]
    (8\label{ridge}) & \emph{Ridge} & 4D & $f(x) = \frac{10000}{\pi^2 N}\sum_{i=1}^N\exp \left( -100 {\sum_{j=1}^d \left(x_j - \frac{i - 1}{N - 1}\right)^2} \right); N = 1000$\\
    \bottomrule
  \end{tabular}
\end{table}

\begin{table}[ht!]
  \caption{Test implementations}
  \label{tab:impl}
  \begin{tabular}{crrcc}
    \toprule
    Implementation & \#CPUs & \#GPUs & Configurations & Details\\
    \midrule
    \emph{Vegas (v5.4.2)} & 2 & 0 & def, vf, tq & official implementation in Cython\\
    \emph{Vegas96} & 96 & 0 & def, vf, tq & Vegas with multiple CPUs\\
    \emph{VegasFlow (v1.3.0)} & 12 & 1 & vf & Tensorflow implementation\\
    \emph{TorchQuad (v0.4.0)} & 12 & 1 & tq & PyTorch implementation\\
    \emph{cuVegas} & 12 & 1 & def, vf, tq & proposed CUDA implementation with Python\\
    \emph{cuVegas8} & 96 & 8 & def, vf, tq & cuVegas with multiple GPUs\\
    \bottomrule
  \end{tabular}
\end{table}

\begin{table}[ht!]
  \caption{Test system configuration}
  \label{tab:hw}
  \begin{tabular}{c|c}
    \toprule
    \multicolumn{2}{c}{System specifications}\\
    \midrule
    CPUs & 2x AMD EPYC 7643, 96 Threads total, base 2.3 GHz, boost 3.6 GHz\\
    Memory & 1 TB DDR4\\
    GPUs & 8x Nvidia A100 SXM4, HBM2e 80 GB, 400W\\
    Software & OS Rocky Linux 8.7, CUDA 11.8, TensorFlow 2.13.0, PyTorch 2.0.1\\
  \bottomrule
\end{tabular}
\end{table}

To ensure a fair and comprehensive evaluation, we adopt a multi-faceted
methodology.
We use three different configurations for our experiments, as detailed in
Table~\ref{tab:config}.
These configurations are designed to align with the fixed parameter choices of the
baseline implementations, thereby facilitating a more equitable comparison.
Particularly TorchQuad and VegasFlow hard-coded configurations have been matched by
adjusting the parameters of Vegas and cuVegas.
The configurations are ``def'' (default) that refers to the default parameter choice of both cuVegas and Vegas,
``vf'' (VegasFlow) that aligns the parameters to those of VegasFlow,
and ``tq'' (TorchQuad) that matches those of TorchQuad.
The running time results are computed as the mean over five runs, following a
warm-up run.

The implementation versions proposed for testing and the related
computational resources are outlined in Table~\ref{tab:impl},
which also shows the parameter configurations tested for each version.
We evaluate cuVegas using a diverse set of test integrands
and two real-world
applications in financial option pricing and quantum physics.
The test integrands, listed in Table~\ref{tab:integrand}, are chosen to
represent a variety of computational complexities and dimensionalities.
The system configuration is reported in Table~\ref{tab:hw}.

\subsection{VegasFill kernel performance scaling}
\begin{figure}[ht!]
    \centering
        \begin{subfigure}{0.3875\linewidth}
    		\includegraphics[width=\linewidth]{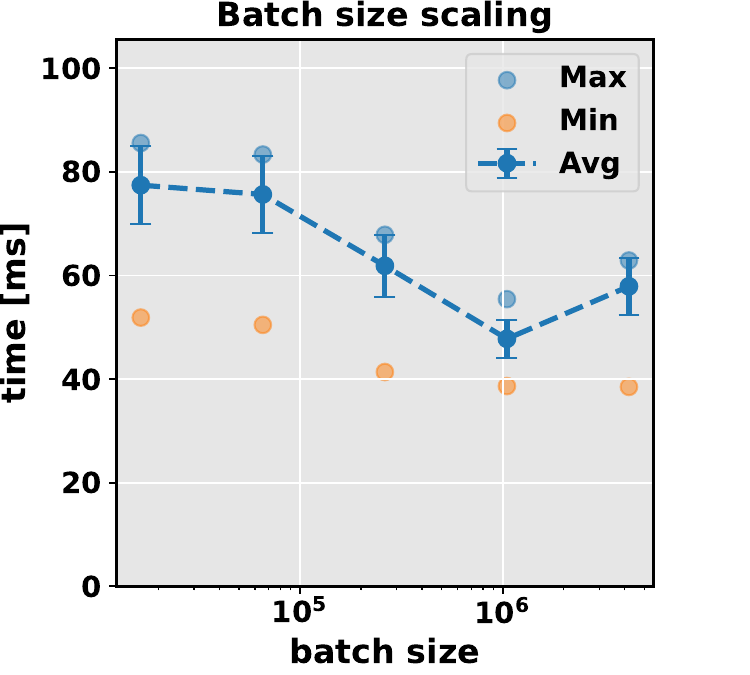}
    		\caption{}
            \label{fig:scaling_a}
        \end{subfigure}
        \begin{subfigure}{0.3875\linewidth}
    		\includegraphics[width=\linewidth]{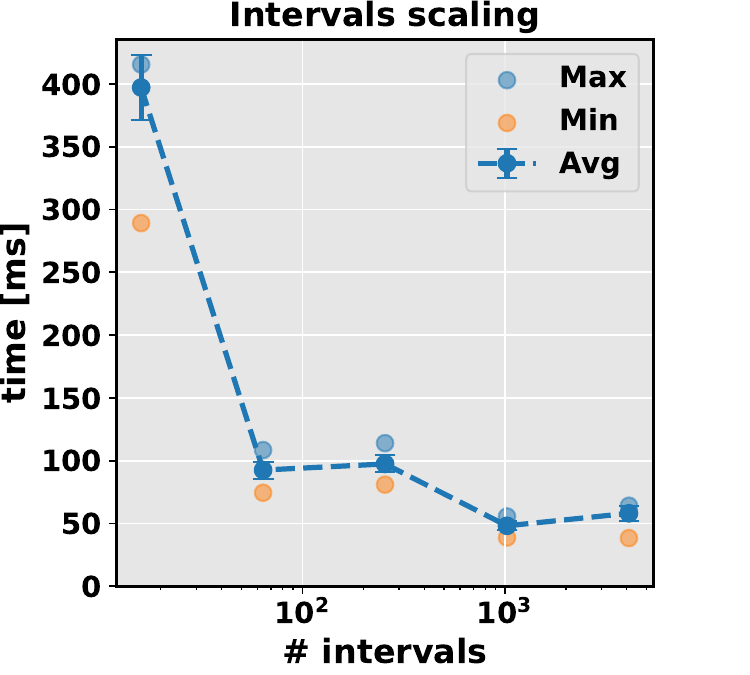}
    		\caption{}
            \label{fig:scaling_b}
        \end{subfigure}\vspace{1em}
        \begin{subfigure}{0.3875\linewidth}
    		\includegraphics[width=\linewidth]{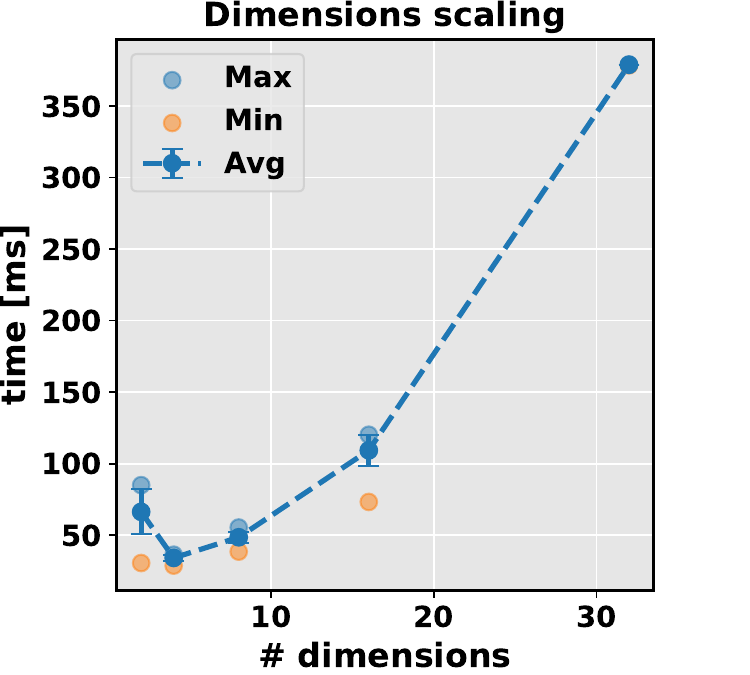}
    		\caption{}
            \label{fig:scaling_c}
        \end{subfigure}
        \begin{subfigure}{0.3875\linewidth}
    		\includegraphics[width=\linewidth]{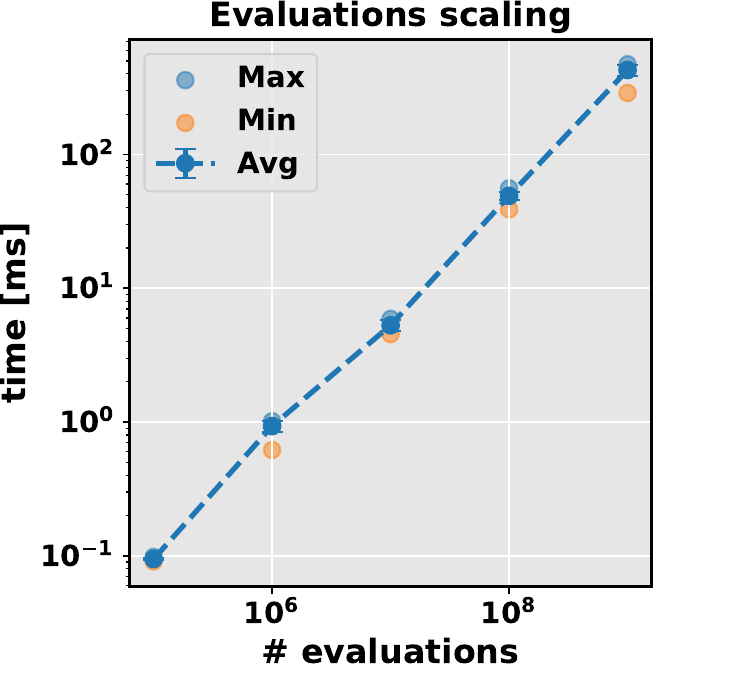}
    		\caption{}
            \label{fig:scaling_d}
        \end{subfigure}
    \caption{VegasFill kernel performance scaling, changing algorithm parameters. The testing parameters are reported in Table~\ref{tab:scaling}. Blue dots represent mean execution time values of the total kernel calls in the program and bars represent standard error. Orange and light blue dots represent minimum and maximum execution time respectively.} \label{fig:scaling}
\end{figure}

We analyzed the execution time performance of the \verb|vegasFill| kernel, when changing the algorithm parameters.
For this testing, the parameter configurations reported in
Table~\ref{tab:scaling} are used, with the \textit{linear} integration
function (\ref{linear}) over 20 iterations.
The default parameters are reported in bold in the table,
and represent the
fixed configuration when changing a single parameter.
The kernel performance is displayed in Figure~\ref{fig:scaling}, showing the
execution time on the $y$-axis when changing the related parameter on the
$x$-axis.

\begin{table}[ht!]
  \caption{Scaling test parameters. The fixed parameters are in reported in bold.}
  \label{tab:scaling}
  \begin{tabular}{c|c|c|c}
    \toprule
    \# evaluations & \# dimensions & \# intervals & batch\_size \\
    \midrule
    \(2e6\) & 2 & 16 & 16384 \\
    \(2e7\) & 4 & 64 & 65536 \\
    \(2e8\) & \textbf{8} & 256 & 262144 \\
    \(2e9\) & 16 & \textbf{1024} & \textbf{1048576} \\
    \textbf{2e10} & 32 & 4096 & 4194304 \\
    \bottomrule
  \end{tabular}
\end{table}

The \verb|batch_size|
parameter determines the number of function evaluations
conducted by a single thread and subsequently influences the grid size,
matching the total number of integrand calls.
As shown in the plot in Figure~\ref{fig:scaling_a}, a lower value has a
detrimental impact on performance, affecting the number of eligible warps and
consequently reducing active warps per scheduler.
On the other hand, higher values involve a tradeoff within a warp, balancing
global memory access randomness and thread serialization for atomic addition.
The value 1,048,576 provides good performance and has been used as default for
all testing scenarios.

The number of intervals in the VEGAS importance sampling map significantly
influences the kernel performance.
As illustrated in the plot in Figure~\ref{fig:scaling_b}, very low values result in a substantial increase
in execution time because all threads atomically sum to the same intervals,
causing thread serialization in the map accumulation.
Conversely, a very high number of intervals decreases the likelihood of
multiple threads writing to the same memory location, yet it leads to
increased global memory reads and diminishes the cache hit ratio due to the
random nature of intervals.
The value 1024 has been chosen as the default, providing a balance between
good performance and compatibility with the algorithm itself, resembling the
original author's choice of 1000 as reported in the VEGAS+ paper \cite{Lepage_2021}
and in the Cython implementation \cite{peter_lepage_2023_8175999}.

As expected, the kernel execution time increases more than linearly with the
number of dimensions in the integrand function, as shown in
Figure~\ref{fig:scaling_c}, due to the additional memory transactions to
global memory.
Additionally, Figure~\ref{fig:scaling_d} shows that increasing the number of
integrand function evaluations leads to an almost linear growth in execution
time, as it primarily involves a linear increase in the number of runs for
each thread.

\subsection{Test integrands benchmark performance}
In order to assess the performance of cuVegas in comparison to TorchQuad and
VegasFlow, performance evaluations were conducted using seven synthetic
benchmarks functions mentioned in Table~\ref{tab:integrand}.
Specifically integrands 1-7 are considered,
excluding Ridge due to its off-scale computational intensity compared to the other functions.
\begin{figure}[ht!]
    \centering
    \includegraphics[width=0.93\linewidth]{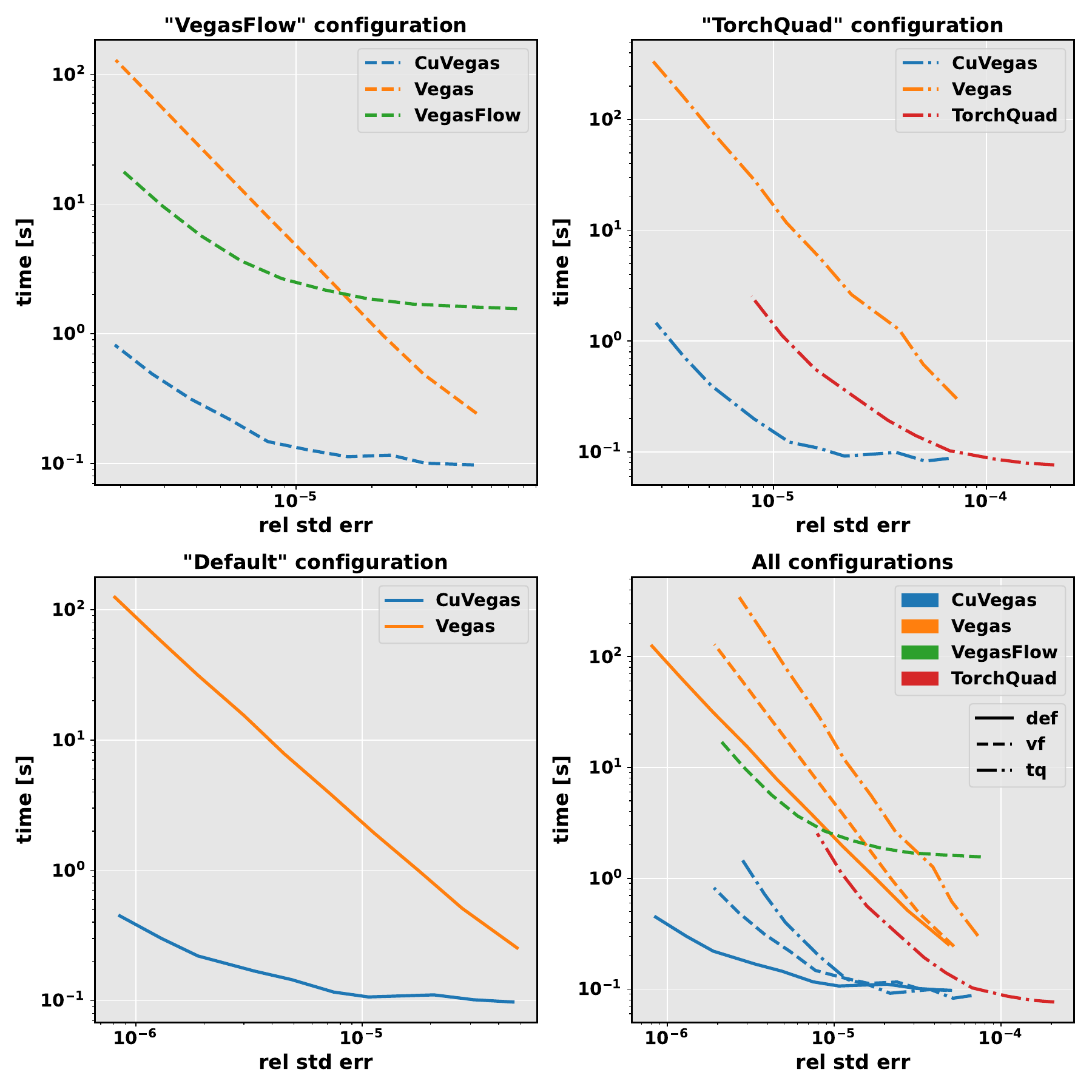}
    \caption{Performance comparison of cuVegas, Vegas, TorchQuad and VegasFlow
    across seven test functions. On the $y$-axis the average wall-clock time is
    plotted against the average relative standard error on the $x$-axis. Axes
    are in log-scale. Lines represent the geometric mean over the seven
    integrands.}
    \label{fig:toy}
\end{figure}

Figure~\ref{fig:toy} depicts the geometric mean of performance across the
seven test integrands.
Each data point corresponds to the mean performance of a given implementation
with a consistent
number of function
evaluations across all integrands.
We show the performance of the different implementations
and configurations reported in Table~\ref{tab:impl},
by showing the results separately for each of the three different configurations,
and then the same results combined in a single plot for comparison.
The geometric mean of the relative standard error on the $x$-axis
is reported against the execution time on the $y$-axis.
It is evident that cuVegas exhibits a promising trend.
Notably, cuVegas consistently demonstrates lower execution times across a
range of relative standard errors, underlining its optimization for rapid
computations.
Speedups in Table~\ref{tab:speedup} have been computed with respect to the
longest run of each version and configuration combination, which result in
almost matching error values.
\begin{table}[ht!]
    \caption{Speedup table presenting the relative performance improvements of cuVegas compared to Vegas, TorchQuad, and VegasFlow on the test integrand experiment.}
    \label{tab:speedup}
    \begin{tabular}{|c|c|c|c|}
        \hline
        \textit{7 Test integrands} & \multicolumn{3}{c|}{Speedup} \\
        \cline{1-4}
        Version/Config & def & vf & tq\\
        \hline
        Vegas & 278.4x & 157.2x & 234.8x \\
        VegasFlow & - & 21.6x & -\\
        TorchQuad & - & - & 6.3x\\
        \hline
    \end{tabular}
\end{table}
However, it is worth mentioning that the TorchQuad result does not match the
errors of the other versions with the same launch parameters, due to an
internal early stopping criterion that stops the computation with a lower
number of function evaluations, and resulting in a higher error compared to
the other versions.
In this case, the speedup is computed comparing the running time of the point
with lowest average error achieved by TorchQuad with the time of the next
point of cuVegas with a lower error.
Considering subsequent points being computed by doubling the number of
n\_evals and assuming a linearly increasing time, the worst-case penalty is 2x.
This puts cuVegas in an unfavorable scenario, but it returns a lower error
while still being faster.
In addition, it is interesting to observe how the ``default'' (``def'') parameter
configuration of Vegas and cuVegas gives better average performance compared
to the ``VegasFlow'' (``vf'') and ``TorchQuad'' (``tq'')
settings across the benchmark functions.
This suggests that the default configuration might be a more sensible
parameter choice on average.

\subsection{Multi-GPU performance scaling}
As described in Section~\ref{sec:multi}, our implementation supports multi-GPU
execution, with the workload distribution among GPUs centered around the
vegasFill part.
Therefore, leveraging multiple GPUs can be advantageous particularly when the
majority of the program execution time is spent on the filling computation,
and the associated data distribution overhead among the GPUs remains
insignificant, in line with the principles of Amdahl's Law.
This advantage becomes notable when dealing with computationally intensive
integrand functions and a substantial number of function evaluations.

\begin{figure}[ht!]
    \centering
    \includegraphics[width=0.5425\linewidth]{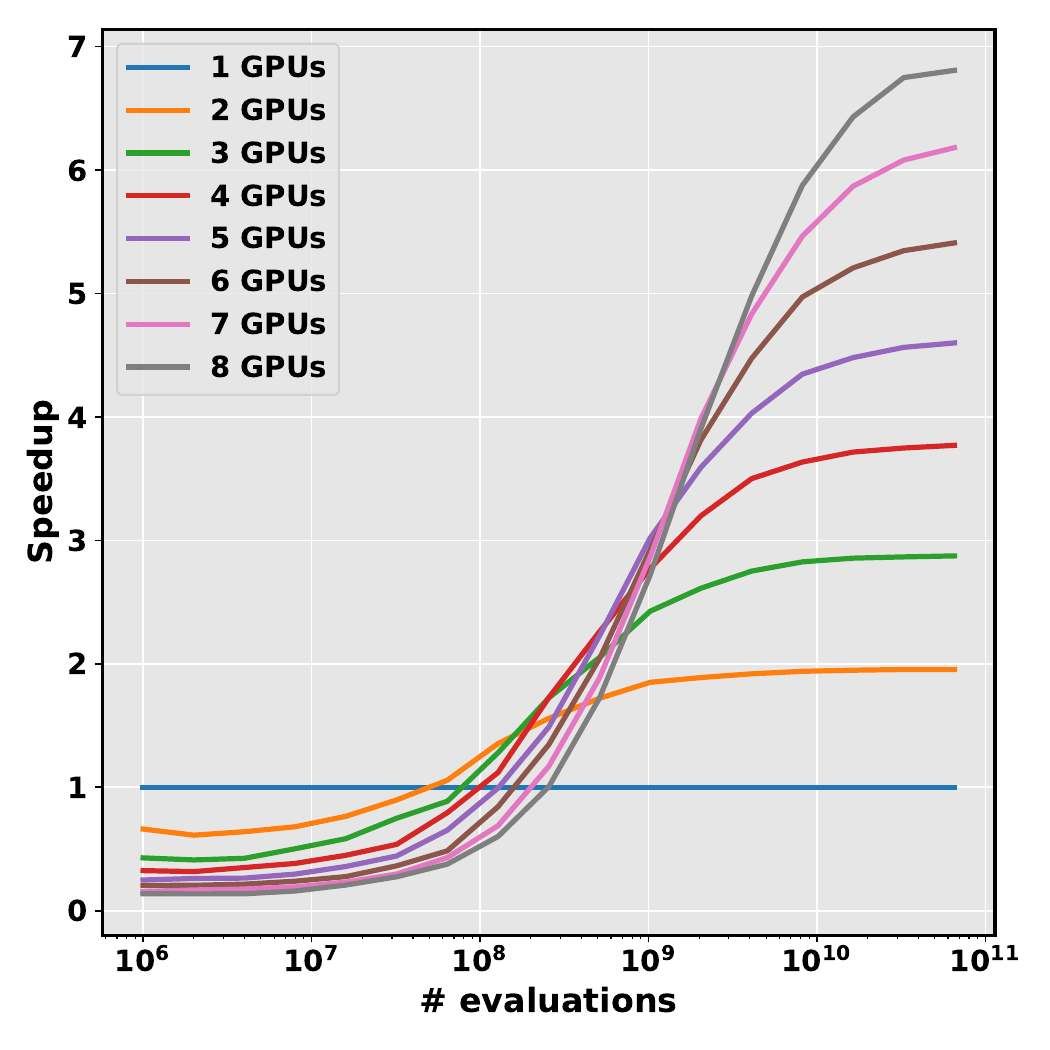}
    \caption{Speedup of multiple GPUs with respect to the single GPU version
    for the \textit{Ridge} integrand, varying the number of function
    evaluations.}
    \label{fig:multi}
\end{figure}

\begin{table}[ht!]
    \caption{Speedup and efficiency table presenting the relative performance improvements of multi-GPU runs compared to a single GPU.}
    \label{tab:multi}
    \begin{tabular}{|c|c|c|}
        \hline
        \# GPUs & Speedup & Efficiency\\
        \hline
        1 & 1.00x & 1.00\\
        2 & 1.95x & 0.98\\
        3 & 2.87x & 0.96\\
        4 & 3.77x & 0.94\\
        5 & 4.60x & 0.92\\
        6 & 5.41x & 0.90\\
        7 & 6.18x & 0.88\\
        8 & 6.81x & 0.85\\
        \hline
    \end{tabular}
\end{table}

As shown by the plot in Figure~\ref{fig:multi}, when considering a
sufficiently computationally intensive integrand, such as the Ridge function
(defined in Table~\ref{tab:integrand}),
using multiple GPUs effectively reduces computation time,
with enough function evaluations.
In Table~\ref{tab:multi}, we report the speedups computed with the highest
number of integrand evaluations considered.
The table also shows the efficiency computed as \(speedup / n_{gpus}\), the
average performance of each GPU with repect to the single one.
The test has been conducted with the CUDA C implementation of the program.
The timing takes into consideration also the CUDA context initialization,
which has been measured around 350 ms for each GPU.
Therefore it is beneficial to enable the multi-GPU execution of the program
when the algorithm running time is considerably larger than the context
initialization, to take advantage of the improved performance in the filling
part when the overhead is less significant.

\subsection{Practical Application Scenarios}
We further extend our experiments to two practical applications scenarios, the
pricing of Asian Options in Finance and the Path Integrals in Quantum
Mechanics.

\subsubsection{Pricing of Asian Options}
Option pricing is a crucial element in financial markets and involves
determining the value of financial derivatives.
The Black-Scholes model, commonly used for European options, can be
computationally demanding when applied to Asian options, largely due to the
high dimensionality involved.
Asian options are a type of path-dependent option where Monte Carlo methods
are an effective computational tool~\cite{Glasserman_2010}.

We consider the following integrand, which corresponds to the computational
representation of an Asian call option:
\begin{equation}
\int e^{-rT} \max(S_{\text{avg}} - K, 0) \, \text{d}x
\end{equation}
where \( e^{-rT} \) is the discount factor, \( S_{\text{avg}} \) is the average asset price over the life of the option, \( K \) is the strike price, \( r \) is the risk-free rate, \( T \) is the time to maturity, and \( \max(S_{\text{avg}} - K, 0) \) represents the payoff of the option. The \( S_{\text{avg}} \) is defined as follows:
\begin{equation}
S_{\text{avg}} = S_0 \exp \left( (r - \frac{\sigma^2}{2})T + \sigma \sqrt{T} \sum_{i=1}^{n} \text{erf}^{-1}(2x_i - 1) \sqrt{2} \right)
\end{equation}
For further details and derivation, we refer to~\cite{Glasserman_2010}.

\begin{figure}[ht!]
    \centering
    \includegraphics[width=0.93\linewidth]{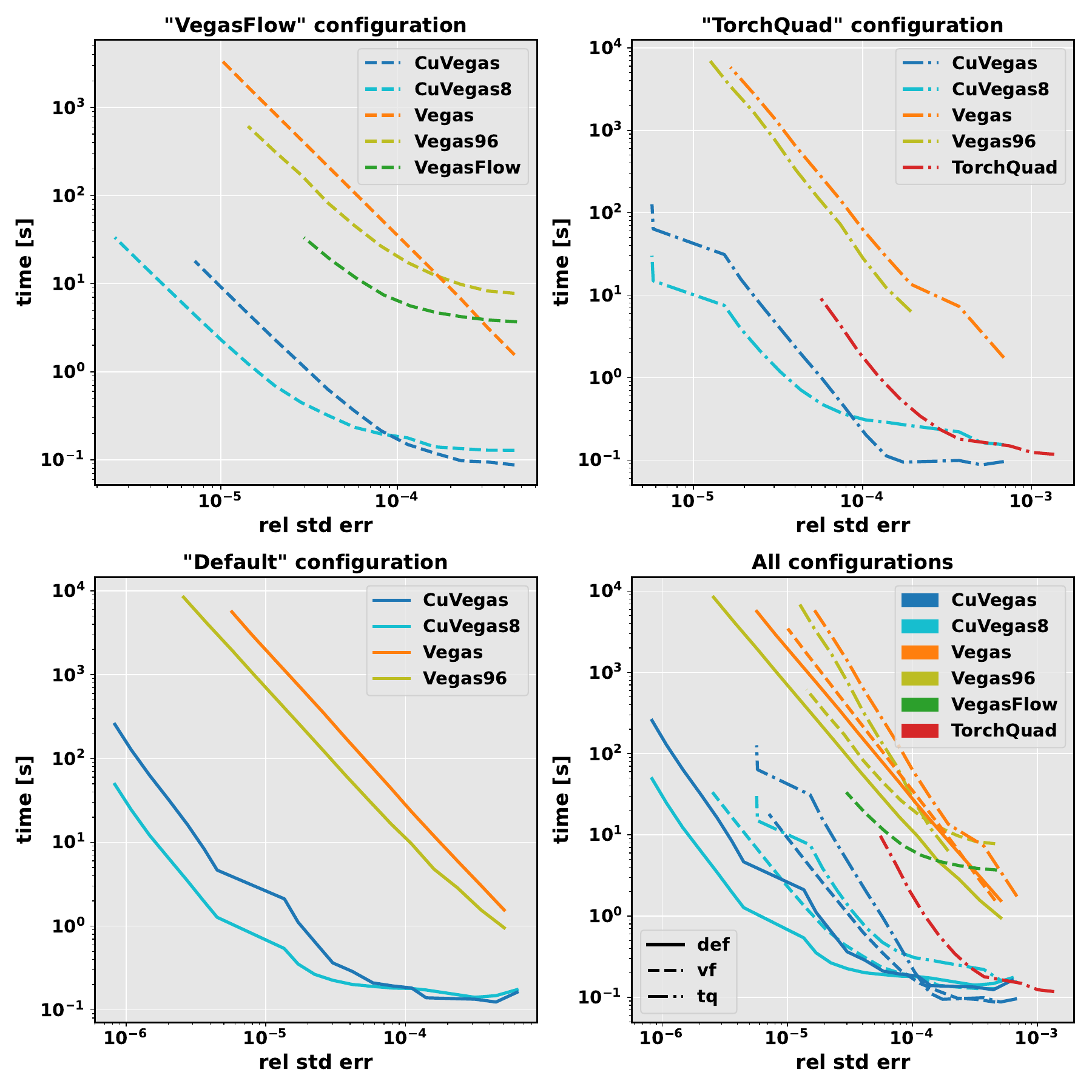}
    \caption{Performance comparison of cuVegas, Vegas, TorchQuad, and VegasFlow on the Asian Option Pricing example integrand. The y-axis plots the average wall-clock time against the average relative standard error on the x-axis, both on log-scale.}
    \label{fig:asian}
\end{figure}

Figure~\ref{fig:asian} showcases the comparative performance of different
integration methods for Asian option pricing.
Speedups in Table~\ref{tab:asian_speedup} are computed in a ``worst case'' manner with respect to cuVegas,
considering the lowest error point achieved by each configuration 
and comparing it to the next computed point of cuVegas with a lower error.
\begin{table}[ht!]
    \caption{Speedup table showing the relative performance improvement of cuVegas
    compared to Vegas, TorchQuad, and VegasFlow in the Asian Option Pricing
    integrand experiment.}
    \label{tab:asian_speedup}
    \begin{tabular}{|c|c|c|c|c|c|c|}
        \hline
        \textit{Asian option} & \multicolumn{6}{c|}{Speedup} \\
        \cline{1-7}
        Config & \multicolumn{2}{c|}{def} & \multicolumn{2}{c|}{vf} & \multicolumn{2}{c|}{tq}\\
        \cline{1-7}
        Version & cuVegas & cuVegas8 & cuVegas & cuVegas8 & cuVegas & cuVegas8\\
        \hline
        Vegas & 1210.2x & 4426.9x & 382.0x & 783.7x & 186.6x & 772.3x\\
        Vegas96 & 261.8x & 1309.3x & 134.1x & 497.6x & 111.6x & 474.5x\\
        VegasFlow & - & - & 27.4x & 74.6x & - & -\\
        TorchQuad & - & - & - & - & 5.0x & 13.8x\\
        \hline
    \end{tabular}
\end{table}
Our method demonstrates superior efficiency across different configurations,
with the ``default'' (``def'') parameter parameter setup 
consistently delivering the best performance. 
This integrand is a computationally intensive function, 
that effectively harnesses the power of an 8-GPU execution, 
resulting in noticeable performance gains. 
It's important to note that both VegasFlow and TorchQuad experience early termination due to memory saturation.

\subsection{Path Integrals in Quantum Mechanics} \label{feynman}
In quantum mechanics, the path integral formulation extends the concept of the
stationary action principle from classical mechanics. In principle, the quantum mechanics of any system (with a classical limit) can
be reduced to a problem in multidimensional integration. In one-dimensional quantum mechanics, the evolution of a position eigenstate
$|x_i\rangle$ from time $t_i$ to time $t_f$ can be computed using a path
integral, We follow the discretization of the path integral procedure described by
Lepage~\cite{LepageQCD} as:
\begin{equation}
\langle x | e^{-\tilde{H}T} | x \rangle \approx A \int_{-\infty}^{\infty} dx_1 \ldots dx_{N-1} e^{-S_{\text{lat}}[x]}
\end{equation}
where the lattice action $S_{\text{lat}}[x]$ is given by:
\begin{equation}
S_{\text{lat}}[x] \equiv \sum_{j=0}^{N-1} \left( \frac{m}{2a}(x_{j+1} - x_j)^2 + aV(x_j) \right),
\end{equation}
with $x_0 = x_N = x$ the fixed endpoints and $a = T/N$ the grid spacing.
We consider a one-dimensional harmonic oscillator, with a potential  $V(x) =
\frac{x^2}{2}$.
The integrand high dimensionality and the presence of peaky features make it a suitable candidate to test the performance of our implementation.

\begin{figure}[ht!]
    \centering
    \includegraphics[width=0.93\linewidth]{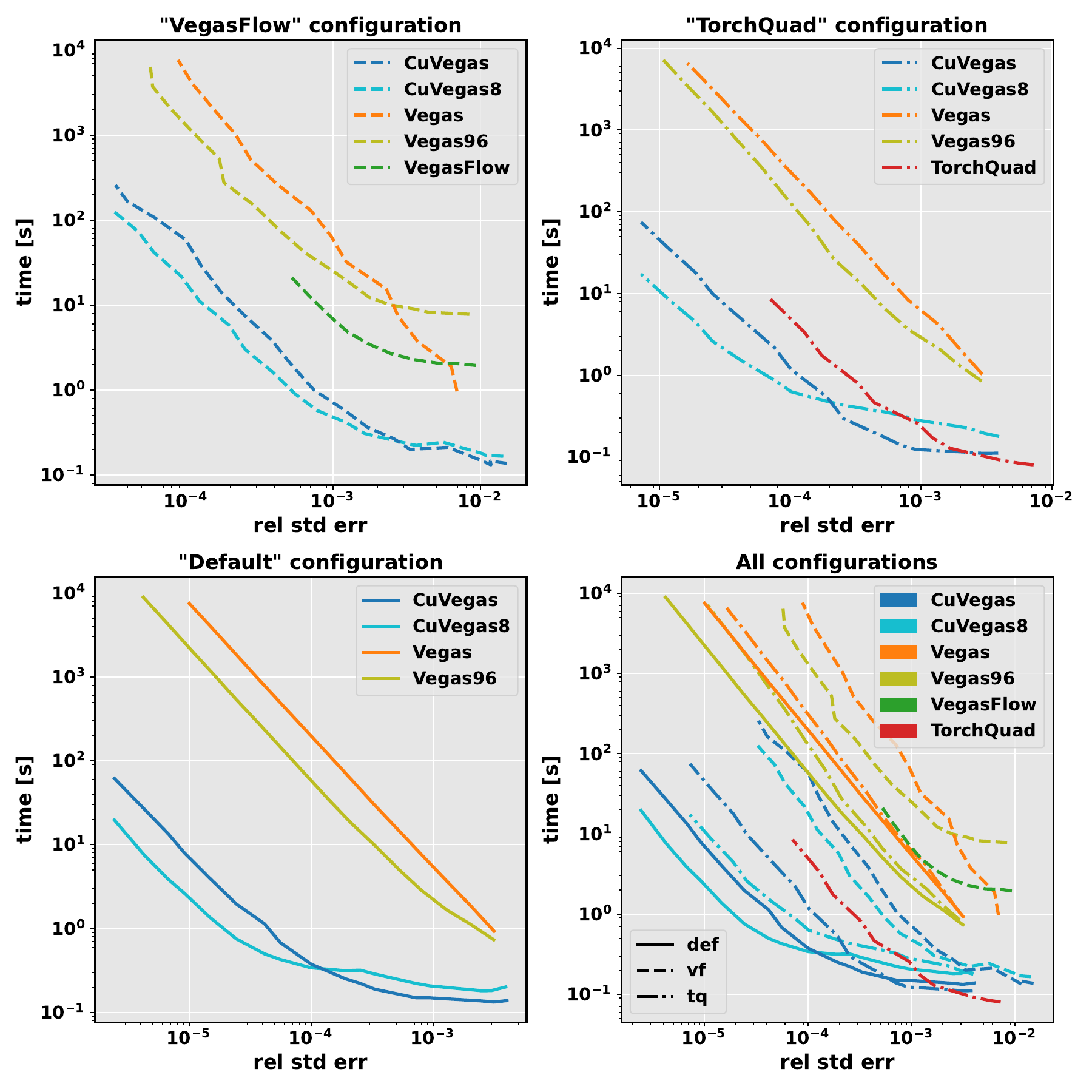}
    \caption{Performance comparison of cuVegas, Vegas, TorchQuad, and VegasFlow on the Feynman Path example integrand. The y-axis shows the average wall-clock time versus the average relative standard error on the x-axis, both in log-scale.} \label{fig:feynman}
\end{figure}

Figure~\ref{fig:feynman} displays a comparative performance analysis of
various integration methods applied to the Path integral.
The speedups, as detailed in Table~\ref{tab:feynman_speedup}, are calculated
similarly to the previous section, with a focus on cuVegas relative
performance.
\begin{table}[ht!]
    \caption{Speedup table showing the performance enhancements of cuVegas compared to Vegas, TorchQuad, and VegasFlow on the Feynman Path integrand experiment.}
    \label{tab:feynman_speedup}
    \begin{tabular}{|c|c|c|c|c|c|c|}
    \hline
    \textit{Feynman path} & \multicolumn{6}{c|}{Speedup} \\
    \cline{1-7}
    Config & \multicolumn{2}{c|}{def} & \multicolumn{2}{c|}{vf} & \multicolumn{2}{c|}{tq}\\
    \cline{1-7}
    Version & cuVegas & cuVegas8 & cuVegas & cuVegas8 & cuVegas & cuVegas8\\
    \hline
    Vegas & 939.9x & 2848.1x & 71.1x & 182.8x & 178.0x & 746.5x\\
    Vegas96 & 145.0x & 452.4x & 41.3x & 93.5x & 96.1x & 413.4x\\
    VegasFlow & - & - & 5.9x & 14.3x & - & -\\
    TorchQuad & - & - & - & - & 1.8x & 5.8x\\
    \hline
    \end{tabular}
\end{table}
Our method, cuVegas, demonstrates improved efficiency in different
configurations compared to the other methods.
This experiment underscores the efficiency of our implementation in handling
integrands with complex features.
It also raises questions about the potential comparative performance of our
method against optimized VEGAS+ implementations for integrals with multiple
peaks or other complex structures.

\subsection{Impact of Adaptive Stratified Sampling}
To assess the specific advantage conferred by adaptive stratified sampling
within the VEGAS algorithm framework, we compare the performance of cuVegas
with the well-optimized m-CUBES 
VEGAS implementation~\cite{DBLP:journals/corr/abs-2202-01753} that we identified as
the best performing implementation of the classical VEGAS algorithm.
This comparison particularly aims to elucidate the efficiency gains achievable
with adaptive stratification, especially for integrands with highly peaked
structures.
For testing, we utilize the m-CUBES implementation from Paterno~\cite{mcubes},
compared to the CUDA C implementation of cuVegas.

\begin{figure}[ht!]
    \centering
    \includegraphics[width=0.93\linewidth]{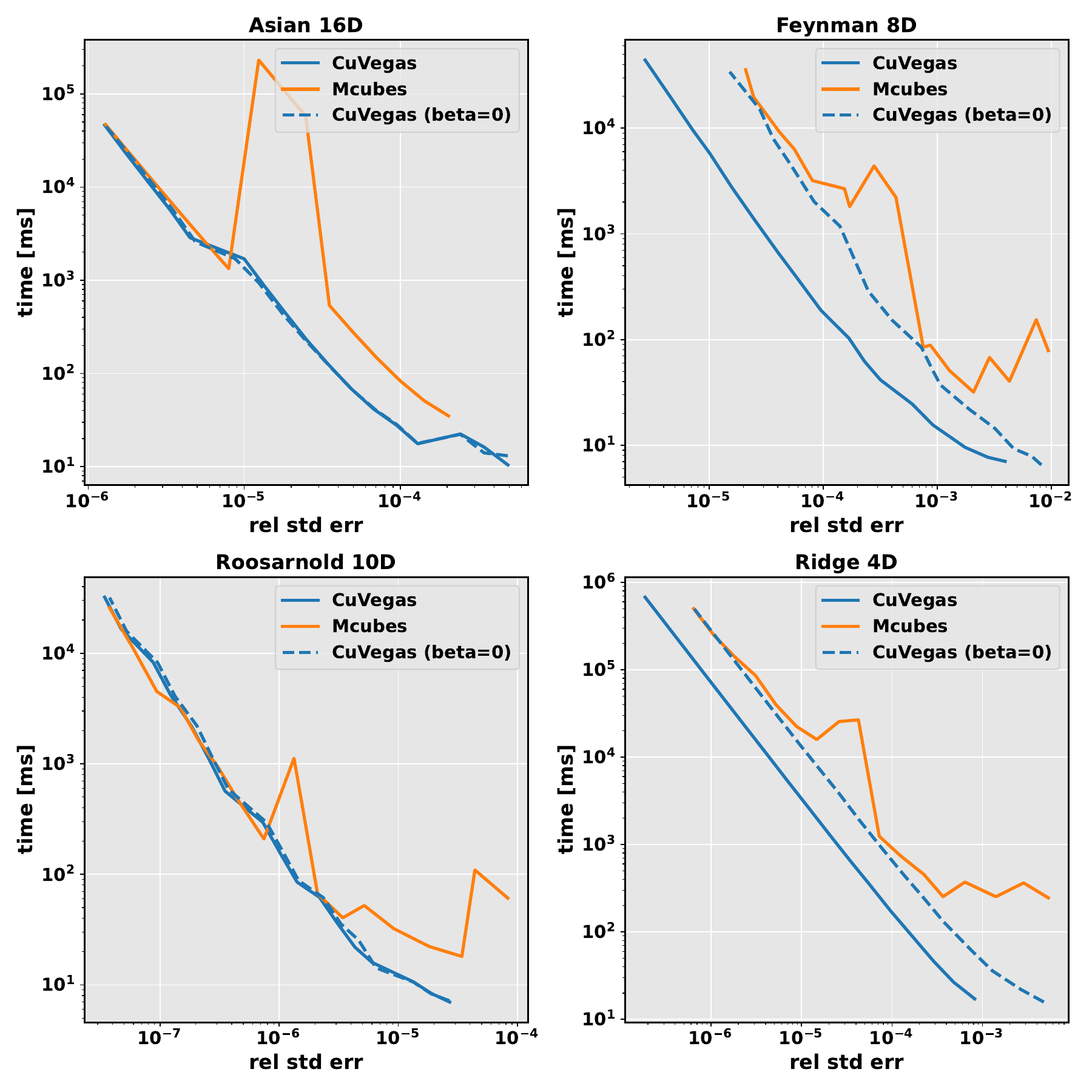}
    \caption{Comparative performance analysis between cuVegas and m-CUBES. On
    the y-axis the average wall-clock time is plotted against the average
    relative standard error on the x-axis. Axes are in log-scale.}
    \label{fig:enhanced}
\end{figure}

Figure~\ref{fig:enhanced} illustrates the results of this head-to-head
assessment.
Here, we configured both cuVegas and m-CUBES with consistent parameters (\(
\alpha = 1.5, n_{\text{intervals}} = 500 \)), corresponding to m-CUBES
default, and we also matched the \(n_{\text{strat}}\) parameter computation to that of m-CUBES,
to ensure an equitable comparison.
cuVegas was tested under two distinct settings to highlight the impact of
adaptive stratification. In detail we enabled adaptive stratification with \( \beta = 0.25 \) (solid line),
and disabled it with \( \beta = 0 \) (dashed line).

The test protocol involves running each algorithm for 20 iterations,
discounting the first 5 iterations to negate initial stabilization
discrepancies in performance recording.
We selected a suite of integrands for evaluation, including those with
pronounced peaks, such as the Ridge and Feynman integrands.

The findings are clear: the performance of cuVegas, when leveraging adaptive
stratification (\( \beta = 0.25 \)), improves markedly for integrands with
significant peak characteristics, confirming the feature value in these
challenging scenarios.
For more uniformly behaving integrands, the performance distinction between
stratification-enabled and -disabled is minimal, with m-CUBES maintaining a
performance parity with cuVegas across a higher number of function
evaluations.

\subsection{Discussion}
Our testing involved a comprehensive set of synthetic benchmark functions and real-world applications, 
including Asian option pricing in finance and Feynman path integrals in quantum physics. 
Using equivalent parameter configurations across implementations ensured fair and consistent comparisons. 
The results showed that cuVegas is consistently faster than both CPU-based and other GPU-based alternatives 
across a variety of computational complexities and dimensionalities.

In real-world scenarios like Feynman path integrals and Asian options, 
cuVegas demonstrated a clear advantage. 
In the context of Asian option pricing, 
our implementation achieved faster computation times and reduced error margins, 
proving its efficacy in financial computations where rapid calculations are necessary for real-time trading and risk management. 
Similarly, Feynman path integrals in quantum physics showed cuVegas's ability to handle complex integrands 
and provide precise calculations, which are crucial for advancements in science and industry.

The default configuration consistently outperformed the VegasFlow and TorchQuad fixed settings, 
indicating that our default parameter setup is generally more effective across different scenarios.
Nevertheless, cuVegas allows users to configure and adapt its parameters, offering flexibility beyond fixed configurations.
Additionally, cuVegas excels at managing complex integrands, especially those with multiple peaks or irregular structures, 
thanks to the adaptive stratified sampling approach of VEGAS+.
In these situations, the program demonstrated its stability and efficiency 
in balancing workload across threads for various integrand scenarios.

The multi-GPU performance scaling experiments further highlight cuVegas's scalability. 
Our implementation shows substantial performance gains when leveraging multiple GPUs, 
especially for computationally intensive tasks. 
This capability is crucial for applications requiring large-scale simulations and high computational throughput.

Despite these significant improvements, cuVegas has some limitations. 
The memory requirements for large-scale integrands can be substantial, 
particularly when scaling across multiple GPUs. 
Additionally, performance gains are primarily observed with complex integrands,
while simpler scenarios might benefit from further optimization,
as the overhead of advanced sampling techniques may not justify the performance gains.

Future work could explore reducing computational costs for simpler integrands 
and expanding support for more diverse hardware configurations, 
including mixed-precision calculations and distributed computing frameworks, 
to extend cuVegas's applicability even further~\cite{BABOULIN20092526}.

\section{Conclusions} \label{concl}
In this work, we introduced cuVegas, a CUDA-based implementation of the Vegas Enhanced algorithm, 
and provided a comprehensive performance analysis against existing CPU-based and GPU-based implementations. 
Our findings clearly demonstrate the significant advantages of leveraging GPU computing for Monte Carlo integration tasks, 
with cuVegas achieving speedups of one to three orders of magnitude over traditional CPU implementations. 
Moreover, when compared to existing Python GPU frameworks~\cite{torchquad_package, Carrazza:2020rdn}, 
cuVegas exhibits a speedup of 2x to 20x, underscoring its superior efficiency, 
outperforming the other implementations across a variety of integrands and dimensionalities.

Against native CUDA C implementations~\cite{mcubes}, cuVegas matches
performance for standard integrands and exhibits a factor of five speedup for
integrands benefiting from adaptive stratified sampling.
This highlights the algorithm's effectiveness in handling complex integration tasks, 
further reinforcing its utility across various applications.

Notably, cuVegas supports multi-GPU setups, enhancing its capability to process complex integrals. 
The inclusion of Python bindings also simplifies the usage of the method, 
making it accessible to a broader audience and facilitating integration with existing scientific workflows.

Extensive evaluation on benchmark functions and real-world scenarios 
like Feynman path integrals in quantum physics and Asian option pricing in financial mathematics, 
indicates that our method's strengths are consistent across various domains. 
This consistency provides confidence in cuVegas's applicability to a wide range of complex integration tasks, 
from financial modeling to quantum physics simulations.

In conclusion, the implementation of the Vegas Enhanced algorithm on GPUs is a promising advancement 
that enhances computational power and applicability across numerous fields. 
By leveraging GPU parallelism, our approach provides a scalable and efficient solution 
for high-dimensional numerical integration challenges.
This capability opens the door to more advanced analyses and practical applications across different fields
and it lays the groundwork for further exploration and development.

\setcounter{footnote}{0}
\renewcommand{\thefootnote}{\alph{footnote}}

\newpage
\printbibliography

@String{Computing = "Computing" }

@String{Computer = "{IEEE} Computer" }

@String{Springer = "Springer-Verlag" }

@article{Kanzaki_2011,
	doi = {10.1140/epjc/s10052-011-1559-8},
  
	url = {https://doi.org/10.1140%2Fepjc%2Fs10052-011-1559-8},
  
	year = 2011,
	month = {feb},
  
	publisher = {Springer Science and Business Media {LLC}
},
  
	volume = {71},
  
	number = {2},
  
	author = {J. Kanzaki},
  
	title = {Monte Carlo integration on {GPU}},
  
	journal = {The European Physical Journal C}
}

@BOOK{test,
   author = "Donald E. Knuth",
   title = "Seminumerical Algorithms",
   volume = 2,
   series = "The Art of Computer Programming",
   publisher = "Addison-Wesley",
   address = "Reading, MA",
   edition = "2nd",
   month = "10~" # jan,
   year = "1981",
}

@article{Lepage_2021,
	doi = {10.1016/j.jcp.2021.110386},
	url = {https://doi.org/10.1016\%2Fj.jcp.2021.110386},
	year = 2021,
	month = {aug},
	publisher = {Elsevier {BV}},
	volume = {439},
	pages = {110386},
	author = {G. Peter Lepage},
	title = {Adaptive multidimensional integration: vegas enhanced},
	journal = {Journal of Computational Physics}
}

@article{PETERLEPAGE1978192,
title = {A new algorithm for adaptive multidimensional integration},
journal = {Journal of Computational Physics},
volume = {27},
number = {2},
pages = {192-203},
year = {1978},
issn = {0021-9991},
doi = {https://doi.org/10.1016/0021-9991(78)90004-9},
url = {https://www.sciencedirect.com/science/article/pii/0021999178900049},
author = {G {Peter Lepage}}
}

@misc{cigar,
  author = {Yongcheng Wu},
  title = {CIGAR},
  year = {2020},
  publisher = {GitHub},
  journal = {GitHub repository},
  howpublished = {\url{https://github.com/ycwu1030/CIGAR}}
}

@misc{cuda,
  author = {Nvidia},
  title = {CUDA Toolkit Documentation},
  year = {2022},
  publisher = {Nvidia},
  url = {https://docs.nvidia.com/cuda/index.html}
}

@misc{vegas,
  author = {Emiliano Tolotti},
  title = {cuVegas},
  year = {2024},
  publisher = {GitHub},
  journal = {GitHub repository},
  howpublished = {\url{https://github.com/emiliantolo/cuvegas}}
}

@software{torchquad_package,
    author       = {Gómez, P. and Hem Toftevaag, H. and Meoni, G.},
    title        = {torchquad: Numerical Integration in Arbitrary Dimensions
      with {PyTorch}, {TensorFlow} \& {JAX} (v0.4.0)},
    year         = 2023,
    publisher    = {Zenodo},
    version      = {v0.4.0},
    doi          = {10.5281/zenodo.8041976},
    url          = {https://doi.org/10.5281/zenodo.8041976}
}

@article{DBLP:journals/corr/abs-2202-01753,
  author       = {Ioannis Sakiotis and
                  Kamesh Arumugam and
                  Marc F. Paterno and
                  Desh Ranjan and
                  Balsa Terzic and
                  Mohammad Zubair},
  title        = {m-CUBES An efficient and portable implementation of multi-dimensional
                  integration for gpus},
  journal      = {CoRR},
  volume       = {abs/2202.01753},
  year         = {2022},
  url          = {https://arxiv.org/abs/2202.01753},
  eprinttype    = {arXiv},
  eprint       = {2202.01753},
  bibsource    = {dblp computer science bibliography, https://dblp.org}
}

@article{Carrazza:2020rdn,
    author = "Carrazza, Stefano and Cruz-Martinez, Juan M.",
    title = "{VegasFlow: accelerating Monte Carlo simulation across multiple hardware platforms}",
    eprint = "2002.12921",
    archivePrefix = "arXiv",
    primaryClass = "physics.comp-ph",
    reportNumber = "TIF-UNIMI-2020-8",
    doi = "10.1016/j.cpc.2020.107376",
    journal = "Comput. Phys. Commun.",
    volume = "254",
    pages = "107376",
    year = "2020"
}

@software{vegasflow_package,
    author       = {Juan Cruz-Martinez and
                    Stefano Carrazza},
    title        = {N3PDF/vegasflow: vegasflow v1.0},
    month        = feb,
    year         = 2020,
    publisher    = {Zenodo},
    version      = {v1.0},
    doi          = {10.5281/zenodo.3691926},
    url          = {https://doi.org/10.5281/zenodo.3691926}
}

@misc{cub,
  author = {Nvidia},
  title = {CUB},
  year = {2023},
  publisher = {Nvidia},
  url = {https://nvlabs.github.io/cub/}
}

@software{peter_lepage_2023_8175999,
  author       = {Peter Lepage},
  title        = {gplepage/vegas: vegas version 5.4.2},
  month        = jul,
  year         = 2023,
  publisher    = {Zenodo},
  version      = {v5.4.2},
  doi          = {10.5281/zenodo.8175999},
  url          = {https://doi.org/10.5281/zenodo.8175999}
}

@article{Feynman,
  title={Space-time approach to non-relativistic quantum mechanics},
  author={Feynman, Richard P},
  journal={Reviews of Modern Physics},
  volume={20},
  number={2},
  pages={367},
  year={1948},
  publisher={APS}
}

@article{Gómez2021,
  doi = {10.21105/joss.03439},
  url = {https://doi.org/10.21105/joss.03439},
  year = {2021},
  publisher = {The Open Journal},
  volume = {6},
  number = {64},
  pages = {3439},
  author = {Pablo Gómez and Håvard Hem Toftevaag and Gabriele Meoni},
  title = {torchquad: Numerical Integration in Arbitrary Dimensions with PyTorch},
  journal = {Journal of Open Source Software}
}

@article{Liu_2001,
  title={Monte Carlo Strategies in Scientific Computing},
  author={J. Liu},
  journal={Springer Series in Statistics},
  year={2001}
}

@article{LepageQCD,
  title={Lattice QCD for Novices},
  author={Lepage, G. Peter},
  journal={arXiv preprint hep-lat/0506036},
  year={2005},
  url={http://arxiv.org/abs/hep-lat/0506036}
}

@article{Kersevan_Richter-Was_2013,
  title={The Monte Carlo event generator AcerMC versions 2.0 to 3.8 with interfaces to PYTHIA 6.4, HERWIG 6.5 and ARIADNE 4.1},
  author={Kersevan, B. P. and Richter-Was, E.},
  journal={Comp. Phys. Comm.},
  volume={184},
  pages={919-985},
  year={2013}
}

@article{Alwall_et_al_2014,
  title={The automated computation of tree-level and next-to-leading order differential cross sections, and their matching to parton shower simulations},
  author={Alwall, J. and others},
  journal={JHEP},
  volume={07},
  pages={079},
  year={2014}
}

@article{Aoyama_et_al_2012,
  title={Complete Tenth-Order QED Contribution to the Muon g -- 2},
  author={Aoyama, T. and Hayakawa, M. and Kinoshita, T. and Nio, M.},
  journal={Phys. Rev. Lett.},
  volume={109},
  pages={111808},
  year={2012}
}

@article{Garberoglio_Harvey_2011,
  title={Path-Integral calculation of the third virial coefficient of quantum gases at low temperatures},
  author={Garberoglio, G. and Harvey, A. H.},
  journal={J. Chem. Phys.},
  volume={134},
  pages={134106},
  year={2011}
}

@article{Campolieti_Makarov_2007,
  title={Pricing Path-Dependent Options on State Dependent Volatility Models with a Bessel Bridge},
  author={Campolieti, G. and Makarov, R.},
  journal={Int. J. Theor. Appl. Finance},
  volume={10},
  pages={51–88},
  year={2007}
}

@article{Serra_Heavens_Melchiorri_2007,
  title={Bayesian Evidence for a cosmological constant using new high-redshift supernova data},
  author={Serra, P. and Heavens, A. and Melchiorri, A.},
  journal={Mon. Not. R. Astron. Soc.},
  volume={379},
  pages={169–175},
  year={2007}
}

@article{Sanders_2014,
  title={Probabilistic model for constraining the Galactic potential using tidal streams},
  author={Sanders, J. L.},
  journal={Mon. Not. R. Astron. Soc.},
  volume={443},
  pages={423–431},
  year={2014}
}

@article{Gultekin_et_al_2009,
  title={The M-$\sigma$ and M-L Relations In Galactic Bulges, and Determinations of their Intrinsic Scatter},
  author={Gultekin, K. and others},
  journal={Astrophys. J.},
  volume={698},
  pages={198—221},
  year={2009}
}

@article{Atay_Hutt_2006,
  title={Neural Fields with Distributed Transmission Speeds and Long-Range Feedback Delays},
  author={Atay, F. M. and Hutt, A.},
  journal={SIAM J. Appl. Dyn. Sys.},
  volume={5},
  pages={670–698},
  year={2006}
}

@article{Dehesa_et_al_2012,
  title={Quantum entanglement in helium},
  author={Dehesa, J. S. and Koga, T. and Yanez, R. J. and Plastino, A. R. and Esquivel, R. O.},
  journal={J. Phys. B: At. Mol. Opt. Phys.},
  volume={45},
  pages={015504},
  year={2012}
}

@misc{pybind11,
   author = {Wenzel Jakob and Jason Rhinelander and Dean Moldovan},
   year = {2023},
   url = {https://github.com/pybind/pybind11},
   title = {pybind11 — Seamless operability between C++11 and Python}
}

@misc{mcubes,
   author = {Marc Paterno},
   year = {2023},
   url = {https://github.com/marcpaterno/gpuintegration},
   title = {Numerical Integration on GPUs}
}

@Manual{cubature,
  title = {cubature: Adaptive Multivariate Integration over Hypercubes},
  author = {Balasubramanian Narasimhan and Steven G. Johnson and Thomas Hahn and Annie Bouvier and Kiên Kiêu},
  year = {2023},
  note = {R package version 2.1.0},
  url = {https://bnaras.github.io/cubature/},
}

@book{10.5555/1538674,
author = {Gough, Brian},
title = {GNU Scientific Library Reference Manual - Third Edition},
year = {2009},
isbn = {0954612078},
publisher = {Network Theory Ltd.},
edition = {3rd}
}

@article{HAHN200578,
title = {Cuba—a library for multidimensional numerical integration},
journal = {Computer Physics Communications},
volume = {168},
number = {2},
pages = {78-95},
year = {2005},
issn = {0010-4655},
doi = {https://doi.org/10.1016/j.cpc.2005.01.010},
url = {https://www.sciencedirect.com/science/article/pii/S0010465505000792},
author = {T. Hahn}
}

@inproceedings{10.1145/2833157.2833162,
author = {Lam, Siu Kwan and Pitrou, Antoine and Seibert, Stanley},
title = {Numba: A LLVM-Based Python JIT Compiler},
year = {2015},
isbn = {9781450340052},
publisher = {Association for Computing Machinery},
address = {New York, NY, USA},
url = {https://doi.org/10.1145/2833157.2833162},
doi = {10.1145/2833157.2833162},
booktitle = {Proceedings of the Second Workshop on the LLVM Compiler Infrastructure in HPC},
articleno = {7},
numpages = {6},
location = {Austin, Texas},
series = {LLVM '15}
}

@book{rubinstein2016simulation,
  title={Simulation and the Monte Carlo method},
  author={Rubinstein, Reuven Y and Kroese, Dirk P},
  year={2016},
  publisher={John Wiley \& Sons}
}

@book{davis2007methods,
  title={Methods of numerical integration},
  author={Davis, Philip J and Rabinowitz, Philip},
  year={2007},
  publisher={Courier Corporation}
}

@book{KalosWhitlock2008,
  title={Monte Carlo Methods},
  author={Kalos, M.H. and Whitlock, P.A.},
  isbn={9783527617401},
  series={Monte Carlo Methods},
  url={https://books.google.it/books?id=IEBjLmmbzDMC},
  year={2008},
  publisher={Wiley}
}

@book{Gelman2013,
  title={Bayesian Data Analysis, Third Edition},
  author={Gelman, A. and Carlin, J.B. and Stern, H.S. and Dunson, D.B. and Vehtari, A. and Rubin, D.B.},
  isbn={9781439840955},
  lccn={2013039507},
  series={Chapman \& Hall/CRC Texts in Statistical Science},
  url={https://books.google.it/books?id=ZXL6AQAAQBAJ},
  year={2013},
  publisher={Taylor \& Francis}
}

@book{Burden2015,
  title={Numerical Analysis},
  author={Burden, R.L. and Faires, J.D. and Burden, A.M.},
  isbn={9781305465350},
  url={https://books.google.it/books?id=9DV-BAAAQBAJ},
  year={2015},
  publisher={Cengage Learning}
}

@book{Glasserman_2010,
  place={New York},
  title={Monte Carlo Methods in financial engineering},
  publisher={Springer},
  author={Glasserman, Paul},
  year={2010}
}

@article{Owens2008,
author = {Owens, John and Houston, Mike and Luebke, David and Green, Simon and Stone, John and Phillips, James},
year = {2008},
month = {05},
pages = {879-899},
title = {GPU computing},
volume = {96},
journal = {Proceedings of the IEEE},
doi = {10.1109/JPROC.2008.917757}
}

@article{BABOULIN20092526,
title = {Accelerating scientific computations with mixed precision algorithms},
journal = {Computer Physics Communications},
volume = {180},
number = {12},
pages = {2526-2533},
year = {2009},
note = {40 YEARS OF CPC: A celebratory issue focused on quality software for high performance, grid and novel computing architectures},
issn = {0010-4655},
doi = {https://doi.org/10.1016/j.cpc.2008.11.005},
url = {https://www.sciencedirect.com/science/article/pii/S0010465508003846},
author = {Marc Baboulin and Alfredo Buttari and Jack Dongarra and Jakub Kurzak and Julie Langou and Julien Langou and Piotr Luszczek and Stanimire Tomov}
}

@article{10.1145/108556.108580,
author = {Favati, Paola and Lotti, Grazia and Romani, Francesco},
title = {Algorithm 691: Improving QUADPACK automatic integration routines},
year = {1991},
issue_date = {June 1991},
publisher = {Association for Computing Machinery},
address = {New York, NY, USA},
volume = {17},
number = {2},
issn = {0098-3500},
url = {https://doi.org/10.1145/108556.108580},
doi = {10.1145/108556.108580},
journal = {ACM Trans. Math. Softw.},
month = {jun},
pages = {218–232},
numpages = {15}
}

@book{10.5555/1964878,
author = {Hwu, Wen-mei W.},
title = {GPU Computing Gems Emerald Edition},
year = {2011},
isbn = {0123849888},
publisher = {Morgan Kaufmann Publishers Inc.},
address = {San Francisco, CA, USA},
edition = {1st}
}

\end{document}